\documentclass[a2paper, 10pt]{wlscirep}
\usepackage[english]{babel}
\usepackage{amsmath, amssymb}
\usepackage{csquotes}
\usepackage{graphicx}
\usepackage{tabularx}
\usepackage{booktabs}
\usepackage{caption}
\usepackage{subcaption}
\usepackage{pgfplotstable}
\usepackage{array}
\usepackage{rotating}
\usepackage{commath}
\usepackage{multirow}
\usepackage{hyperref}
\usepackage{float}
\usepackage[T1]{fontenc}
\usepackage{appendix}
\usepackage[referable]{threeparttablex}
\usepackage[utf8]{inputenc}

\newcommand{\covid}{COVID-19}
\newcommand{\Harmful}{Potentially disinformative}
\newcommand{\harmful}{potentially disinformative}
\newcommand{\Rcluster}{Infrequent posters}
\newcommand{\rcluster}{infrequent posters}
\newcommand{\Scluster}{Intermediate type}
\newcommand{\scluster}{intermediate type}
\newcommand{\pvalue}{$p$-value}

\DeclareCaptionListFormat{tabwithname}{\tablename~#2}
\newcommand*{\supplref}[1]{\hyperref[{#1}]{Supplementary \autoref*{#1}}}

\renewcommand{\appendixpagename}{Supplementary Information for: The connection between the spread of misinformation, time of day, and individual user activity patterns}

\title{The connection between the spread of misinformation, time of day, and individual user activity patterns}

\author[1,*]{Elisabeth Stockinger}
\author[2]{Riccardo Gallotti}
\author[1]{Carina I. Hausladen}

\affil[1]{Computational Social Science, Department of Humanities, Social and Political Sciences, ETH Zurich, Zurich, 8092, Switzerland}
\affil[2]{Complex Human Behaviour Lab, Fondazione Bruno Kessler, Trento, 38123, Italy}

\affil[*]{elisabeth.stockinger@live.at}


\begin{abstract}
Social media manipulation poses a significant threat to cognitive autonomy and unbiased opinion formation. 
Prior literature explored the relationship between online activity and emotional state, cognitive resources, sunlight and weather. 
However, a limited understanding exists regarding the role of time of day in content spread and the impact of user activity patterns on susceptibility to mis- and disinformation. 
This work uncovers a strong correlation between user activity patterns and the tendency to spread manipulated content. 
Through quantitative analysis of Twitter data, we examine how user activity throughout the day aligns with chronotypical archetypes. 
Evening types exhibit a significantly higher inclination towards spreading potentially manipulated content, which is generally more likely between 2:30 AM and 4:15 AM. 
This knowledge can become crucial for developing targeted interventions and strategies that mitigate misinformation spread by addressing vulnerable periods and user groups more susceptible to manipulation.
\end{abstract}
\begin{document}

\flushbottom
\maketitle

\thispagestyle{empty}

\section*{Introduction}

Collective intelligence and democracy rest on the shoulders of public free access to unbiased and diverse information~\cite{Mann2017, Kuklinski2000}.
Social media blurs the borders between news creation, consumption, and distribution~\cite{Kim2021}, as well as between personal communication, announcements from individuals, fiction, and advertisement. Along with the optimization criteria employed in recommendation algorithms~\cite{Diakopoulos2019, Nechushtai2019} and network structures, this contributes to the creation and spread of mis- and disinformation online~\cite{Kim2021}, to political manipulation~\cite{Lin2019, Spaiser2017, Quattrociocchi2011, Saurwein2020, Susser2019}, a collapse of content diversity~\cite{Lazer2015, Bakshy2015, Heitz2022} and political polarisation~\cite{VanBavel2021}. 

This leaves the responsibility to distinguish between the content types and discern truth from deception to the user. However, our ability to scrutinise new information for its reliability depends on the individual's internal state. Cognitive resources and one's thinking style~\cite{Bronstein2019, Bago2020, Pennycook2019, Pennycook2021, Martel2020, Lyons2021, Mosleh2021, Roozenbeek2020, Imhoff2022, Scherer2021, Evans2003, Effron2020, Kahan2017, KnoblochWesterwick2020, Drummond2017, Kahan2017, Kahan2012, Ballarini2017}, as well as emotional state~\cite{Forgas2019, Forgas2008, Weeks2015, MacKuen2010, Martel2020}, have been explored extensively in this regard with diverging results. Other influential factors include cognitive biases and prior beliefs~\cite{Pronin2022, Kim2021, VanBavel2018, Dreyfuss2017, Kahan2017, Lewandowsky2012, SwireThompson2020}.

These factors are not constant but exhibit regular cyclical behaviours with periods ranging from hours to seasons~\cite{Dzogang2017, Golder2011, Lampos2013, Murnane2015, Gleasure2020} and depend on external factors such as light exposure~\cite{Kent2009, Leypunskiy2018, Roenneberg2007}, atmospheric conditions~\cite{Baylis2018, Stevens2021}, social interactions~\cite{Roenneberg2007}, or the device used to access social media~\cite{Murthy2015, Groshek2016, Dunaway2021, Honma2022}. 
Beyond these \emph{zeitgebers}, there are inter-individual differences, particularly affecting circadian process timings. A process is referred to as circadian if it recurs naturally on a twenty-four-hour cycle, and as diurnal if there is a recurrence which may or may not be endogenous. These differences include diverging phase preferences known as chronotypes~\cite{Duarte2021} (so-called \enquote{early birds} or \enquote{night owls}).
In the absence of disruptions to one's natural rhythms, chronotypes perform better at optimal times with \enquote{evening types} achieving better results in the evening, and \enquote{morning types} in the morning~\cite{Taillard2021}. Depending on environmental or social constraints, sleep and activity timings may be out of phase with one's internal circadian time, leading to deterioration in cognitive performance such as attention, memory, or decision-making capacity~\cite{Taillard2021} as well as reflective thinking~\cite{Oyebode2021}. Finally, sleep loss itself has long-reaching effects such as reductions in altruistic behaviour~\cite{Simon2022}.

In an additional layer of complexity, social media are dynamic: They follow human circadian or diurnal rhythms, \cite{Kates2021, Dzogang2018} or the weekday-weekend rhythms \cite{Mayor2021, Dzogang2017}. The timing of a Twitter post is an essential factor in its spread and popularity \cite{Gleasure2020}. 
Clocktime and sunrise/sunset hours have distinct impact on tweeting activity \cite{Dzogang2017}. 

Despite all efforts to mitigate mis- and disinformation \cite{Munson2013, Park2009, Jeon2021, Gillani2018},
they continue to be a substantial problem, even rising in importance with geopolitical (e.g.~\cite{Zawadzki2022}) and epidemiological developments (e.g.~\cite{Roozenbeek2020}). Especially the global \covid{} pandemic has invited a new wave of conspiracy theories~\cite{Roozenbeek2020}, with up to a third of the population believing \covid{} to have been bio-engineered~\cite{Roozenbeek2020}. As an event with drastic and synchronous impact across a major part of the population, the pandemic may have contributed fundamentally to polarisation~\cite{Condie2021}.

\begin{figure*}[bpt]
    \centering
    \caption{Factors influencing the spread of mis- and disinformation. Our study examines the impact of daylight, time of day, human diurnal activity, affiliation to chronotype, and the \covid{} pandemic.}
    \label{fig:model_vars}
    \includegraphics[width=.9\linewidth]{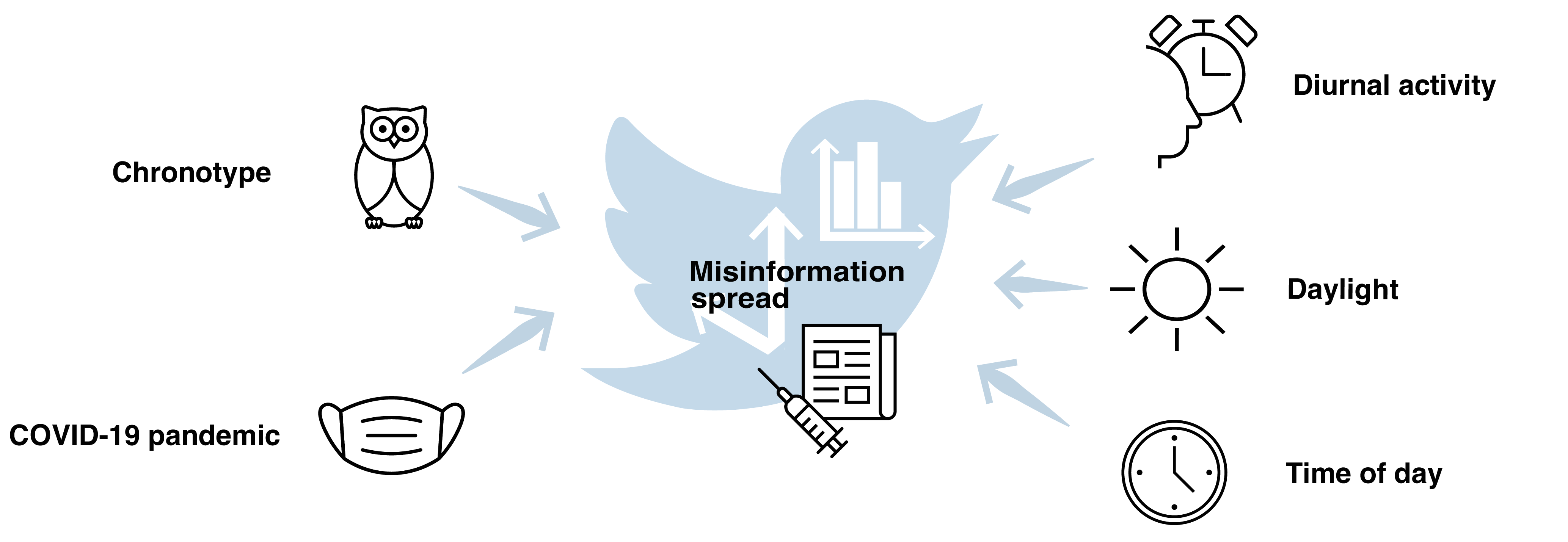}
\end{figure*}

We contribute to this literature by investigating mis- and disinformation about social media --- well summarised, for example, by Tucker \emph{et al.} (2018)~\cite{Tucker2018} --- with an analysis of the interaction effects between temporal rhythms of disinformation and social media usage in the context of \covid{}. 
Specifically, we aim to answer the research question of how the spread of mis- and disinformation on Twitter varies throughout the day. Additionally, we explore whether there are individual differences in users' propensity to spread mis- and disinformation on Twitter based on their typical diurnal activity patterns, both during the day and as a general inclination. \autoref{fig:model_vars} visualises these connections.

\section*{Results}

We analysed a secondary Twitter dataset~\cite{Gallotti2020} relating to the \covid{} pandemic. 
Only tweets containing a link to a different website were included in the dataset. Those tweets were classified into eight categories, also called content types, according to an expert rating of the reliability of the link's domain. We further grouped the categories into those potentially designed to be disinformative, and those that are unlikely to be so.
The categories alongside their user activity statistics are detailed in \supplref{tab:content_categories}. See the \nameref{sec:methods} for further details.

\subsection*{Four prototypical activity patterns}
Our analysis focuses on the individual usage patterns on Twitter and their daily fluctuations. To that end, we first compute the average posting activity of each user per day, including Tweets, Retweets, and Replies. We then use hierarchical clustering to cluster the average posting activity curves.
The analysis reveals the presence of three distinct clusters with unique patterns of posting activity. Users with low post rates ($<240$ posts across the time span under analysis) are separated into a fourth cluster. While this paper focuses on Tweets originating from Italy, we conducted the same analysis for Tweets originating from Germany and found these prototypical activity patterns to hold across the two countries (\ref{sec:suppl:Germany}).

\begin{figure*}
    \centering
    \caption{Smoothed diurnal activity and ratio of \harmful{} content posted per cluster. For each cluster, the two highest peaks of activity and ratio are stressed and annotated with time of occurrence. The shaded area in panel (b) stresses the closeness of peak activity after waking times across the clusters.\label{fig:fourier_activity}}
    \begin{subfigure}[b]{\textwidth}
        \centering
        \includegraphics[height=1.2em]{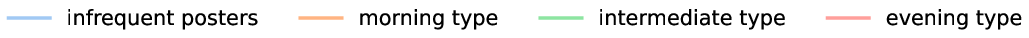}\\
        \includegraphics[height=2em]{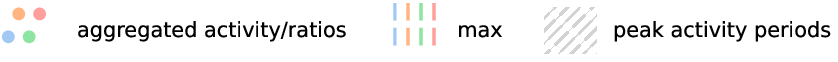}
    \end{subfigure}
    \begin{subfigure}[b]{.5\textwidth}
        \includegraphics[width=\columnwidth]{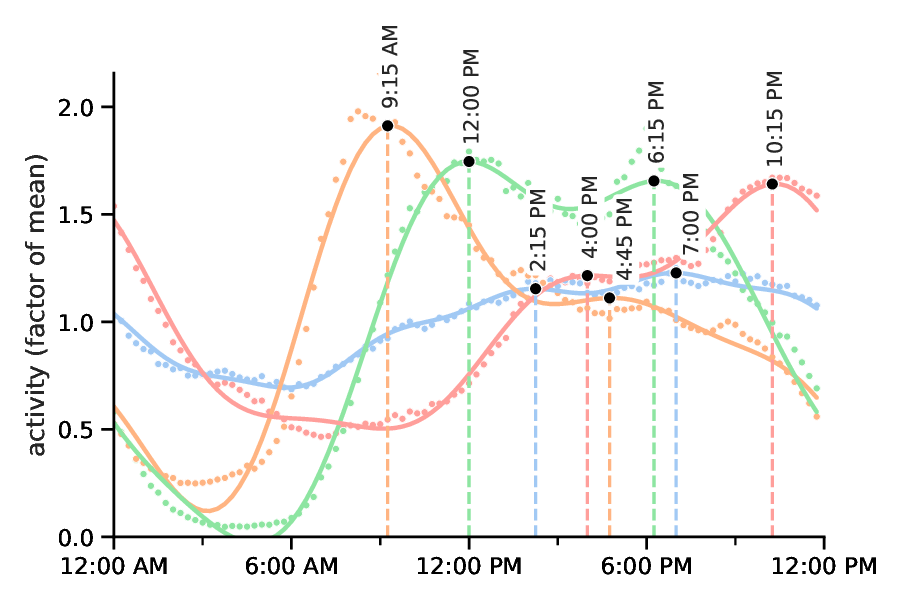} %
        \caption{activity by time of day\label{fig:fourier_activity_clocktime}}
    \end{subfigure}%
    \begin{subfigure}[b]{.5\textwidth}
        \includegraphics[width=\columnwidth]{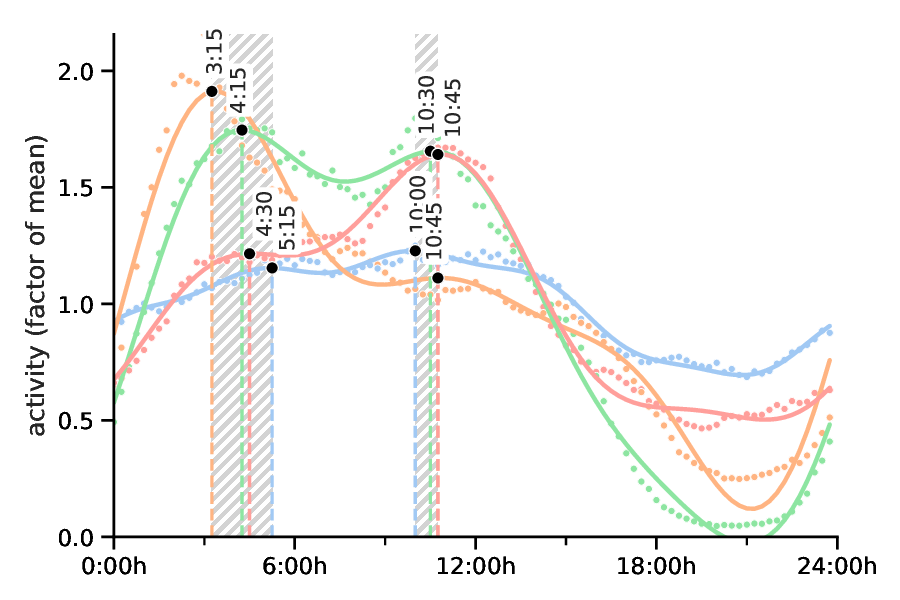}
        \caption{activity by  waking time\label{fig:fourier_activity_waking}}
    \end{subfigure}
    \begin{subfigure}[b]{.5\textwidth}
        \includegraphics[width=\columnwidth]{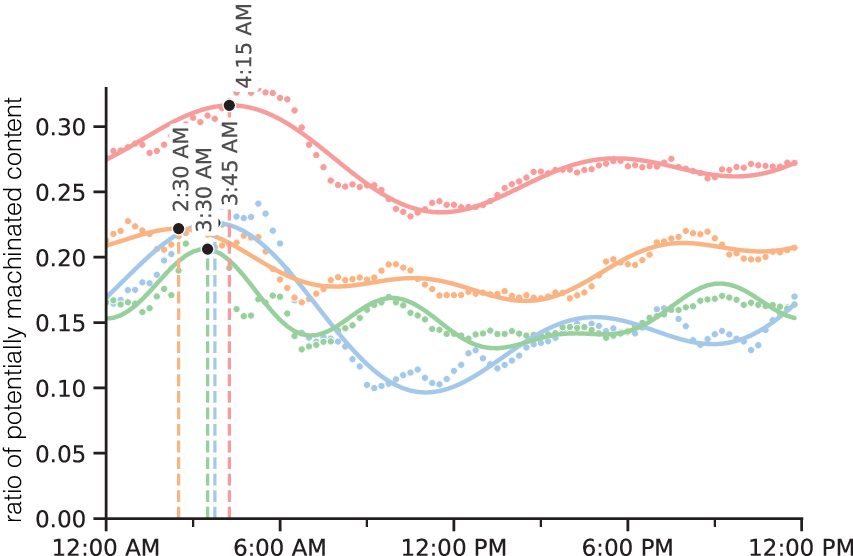} %
        \caption{ratio by time of day\label{fig:fourier_ratios_clocktime}}
    \end{subfigure}%
    \begin{subfigure}[b]{.5\textwidth}
        \includegraphics[width=\columnwidth]{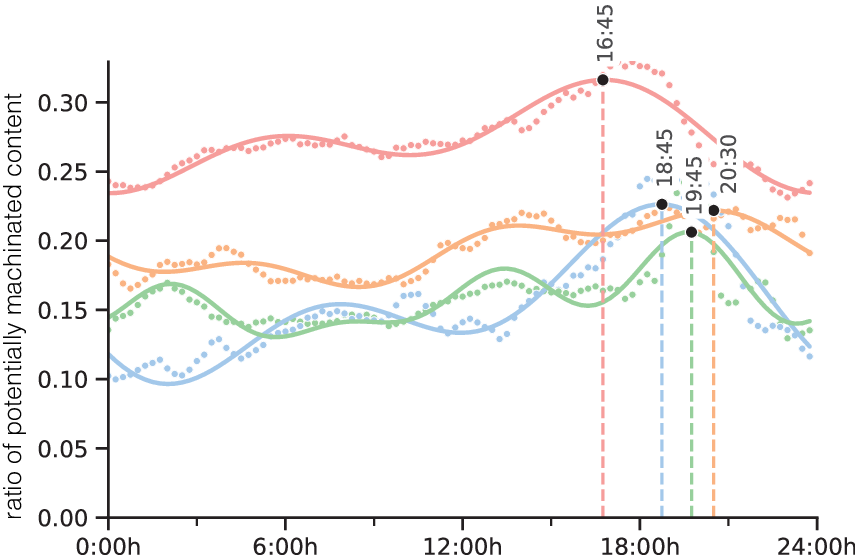} %
        \caption{ratio by waking time \label{fig:fourier_ratios_waking}}
    \end{subfigure}
\end{figure*}

Figure \ref{fig:fourier_activity_clocktime} illustrates the activity patterns of the four clusters throughout the day. 
It indicates the smoothed posting activity for each cluster along with the two largest respective peaks, given in detail in \supplref{tab:activity_ratio_stats}.
We refer to the clusters as morning, evening, and intermediate type posters, named after their respective peak activity times, as well as infrequent type posters.
Generally, user activity follows a bimodal distribution (\supplref{tab:dip_test_activity} shows the Dip-test results rejecting single-modality). 
The orange curve represents \emph{morning types}, with the curve reaching its maximum in the morning at 9:15 AM at twice the average value.
In contrast, \emph{evening types}, displayed in red, exhibit their highest activity at around 10:15 PM.
\emph{Intermediate types}, represented by the green curve, display two nearly identical peaks in size, with the highest peak occurring at noon.
The \emph{infrequent type} group, represented by the blue curve, showed consistent activity levels throughout the day. This cluster groups users who have contributed only a few posts to the dataset, irrespective of activity distribution throughout the day. As a result, the cluster likely includes users from various chronotypes. Their activity patterns may average out over the course of the day, resulting in a relatively flat curve.

We extrapolate from the users' diurnal activity patterns on Twitter to sleeping and waking cycles, which can vary significantly between clusters.
We consider the 16 continuous hours of highest aggregated activity to be a user's average waking time. Consequently, we consider activity outside of this interval to represent prolonged wakefulness, where the user is active despite it being a time of habitual rest. 
A formal definition is given in \autoref{eq:heightened_activity}. Onset and end values of increased activity for each cluster are listed in \supplref{tab:dip_test_activity} (\enquote{heightened activity}).

\autoref{fig:fourier_activity_waking} aligns the clusters' activity by waking time. From this perspective, the diurnal activity curves for each cluster show remarkable similarities. The peaks for all clusters fall within a distinct time window (shaded in grey in the figure). 
The first peak of activity occurs within 3:15 and 5:15 hours after waking within a window of 2 hours. The second peaks of activity occur within 10 to 10:45 hours after waking in a strikingly short window of only 45 minutes.
The sizes of the respective peaks in activity seem to be as much of a differentiating characteristic for each chronotype as the time of occurrence of peak activity.
The activity valleys across clusters are similarly close, occurring around 3 hours before waking (\supplref{tab:activity_ratio_stats}).

\subsection*{\emph{Evening types} spread most \harmful{} content, \emph{\rcluster{}} the least.}
The clusters show distinct features beyond their typical activity patterns. In particular, we find a significant association between content type and cluster affiliation ($\chi^2=15,330.98$, \pvalue$<.001$). 

\autoref{fig:fourier_ratios_clocktime} shows the fluctuation of the ratio of \harmful{} content throughout the day.
Notably, ratios for \emph{evening types}, ranging between $.23$ and $.32$, are consistently elevated as compared to the other clusters (see \autoref{tab:mannwhitneyu_contentypes} for statistical significance and 
\supplref{tab:content_categories} for the distinct variation in ratios of content types spread by cluster).

\begin{table*}
\caption{One-sided Mann-Whitney U test indicating whether the distribution of ratios of \harmful{} content throughout the day (see \autoref{fig:fourier_activity_clocktime}) underlying one cluster (rows) is smaller than that of another cluster (columns).}
\label{tab:mannwhitneyu_contentypes}
\begin{tabularx}{\textwidth}{l*{6}{>{\raggedleft\arraybackslash}X}}
\toprule
 & \multicolumn{2}{c}{morning} & \multicolumn{2}{c}{intermediate} & \multicolumn{2}{c}{evening} \\
 \cmidrule(lr){2-3}\cmidrule(lr){4-5}\cmidrule(lr){6-7}
& $U$ & \pvalue{} & $U$ & \pvalue{} & $U$ & \pvalue{} \\
\midrule
infrequent & 1,705 & \bfseries 2.4e-14 & 3,447 & \bfseries 1.3e-03 & 82 & \bfseries 3.3e-32 \\
morning & - & - & 8,489 & 1.0e+00 & 0 & \bfseries 2.6e-33 \\
intermediate & - & - & - & - & 46 & \bfseries 1.1e-32 \\
\bottomrule
\end{tabularx}

\end{table*}

\emph{\Rcluster{}} exhibit the lowest ratios of \harmful{} content.
This can again be explained by the definition of this cluster as grouping users for whom there are few posts in the dataset, as total posting activity is positively correlated with the dissemination of \harmful{} content.

There is a positive correlation between the amount of posts per user in the dataset and the ratio of \harmful{} content across all users ($\rho = .199$, \pvalue $<.001$) as well as within each cluster (\autoref{tab:total_posts_corr}). 

\begin{table*}
    \centering\caption{Correlation tables in between diurnal and total posting activity and \harmful{} content activity.}
    \begin{subtable}[t]{\linewidth}
        \centering
        \caption{Spearman's rank correlation coefficient and corresponding \pvalue{} comparing a user's total number of posts with ratios of \harmful{} content.}
        \label{tab:total_posts_corr}
        \begin{tabular}{l *{2}{r}}
\toprule

& \multicolumn{2}{c}{\Harmful{}}  \\
 \cmidrule(lr){2-3}
  
& Spearman's $\rho$& \pvalue{} \\
\midrule

infrequent & 0.162 & \bfseries 1.2e-02 \\
morning & 0.185 & \bfseries 1.8e-10 \\
intermediate & 0.068 & \bfseries 2.2e-02 \\
evening & 0.182 & \bfseries 3.2e-10  \\
\midrule
total & 0.199 & \bfseries 2.7e-22 \\
\bottomrule
\end{tabular}
    \end{subtable}
    \begin{subtable}[t]{\linewidth}
        \centering
        \caption{Spearman's rank correlation coefficient and corresponding \pvalue{} comparing a user's (a) raw aggregated activity level (\enquote{coarse}) and (b) the smoothed set of diurnal user activity $ \{ A_{(t,c)} \}_{(t,c) \in T x C}$ (\enquote{smooth}) at different time points in a day with the ratios of politically biased information, fake or hoax news, and conspiracy or junk science as well as all \harmful{} content, negatively weighted by the author's total number of posts. In row \enquote{smooth}, the smoothed set of \harmful{} content $ \{ R_{(t,c)} \}_{(t,c) \in T x C}$ was used to compute the correlation coefficient.}
        \label{tab:diurnal_activity_corr}
        \begin{tabularx}{\linewidth}{p{.5em} X *{8}{r}}
\toprule

& & \multicolumn{2}{c}{Political} 
  & \multicolumn{2}{c}{Fake or hoax} 
  & \multicolumn{2}{c}{Conspiracy \& junk science}
  & \multicolumn{2}{c}{\Harmful{}} \\
  
 \cmidrule(lr){3-4}
 \cmidrule(lr){5-6}
 \cmidrule(lr){7-8}
 \cmidrule(lr){9-10}
 
& & Spearman's $\rho$& \pvalue{} 
  & Spearman's $\rho$& \pvalue{} 
  & Spearman's $\rho$& \pvalue{} 
  & Spearman's $\rho$& \pvalue{} \\

\midrule
\multirow{5}{*}{\rotatebox[origin=c]{90}{\small coarse}}
& infrequent & -0.208 & \bfseries 4.2e-02 & -0.529 & \bfseries 2.9e-08  & 0.263 & \bfseries 9.6e-03 & -0.307 & \bfseries 2.3e-03 \\
& morning & -0.622 & \bfseries 1.3e-11 & -0.514 & \bfseries 8.3e-08 & 0.419 & \bfseries 2.1e-05 & -0.507 & \bfseries 1.4e-07 \\
& intermediate & -0.626 & \bfseries 9.1e-12 & -0.016 & 8.8e-01 & 0.182 & 7.7e-02 & -0.332 & \bfseries 9.4e-04 \\
& evening & -0.097 & 3.5e-01  & 0.034 & 7.4e-01 & 0.157 & 1.3e-01 & 0.045 & 6.6e-01 \\
\cmidrule{2-10}
& total & -0.209 & \bfseries 4.1e-02 & -0.529 & \bfseries 3.0e-08 & 0.263 & \bfseries 9.6e-03 & -0.308 & \bfseries 2.3e-03  \\

\midrule
\multirow{5}{*}{\rotatebox[origin=c]{90}{\small smooth}}
& infrequent & -0.188 & 6.7e-02 & -0.550 & \bfseries 6.3e-09 & 0.274 & \bfseries 7.0e-03 & -0.496 & \bfseries 2.8e-07 \\
& morning & -0.599 & \bfseries 1.1e-10 & -0.542 & \bfseries 1.2e-08 & 0.409 & \bfseries 3.5e-05 & -0.781 & \bfseries 6.5e-21 \\
& intermediate  & -0.604 & \bfseries 7.1e-11 & -0.037 & 7.2e-01  & 0.199 & 5.1e-02 & -0.598 & \bfseries 1.2e-10\\
& evening & -0.030 & 7.7e-01 & 0.058 & 5.7e-01  & 0.089 & 3.9e-01 & -0.086 & 4.1e-01\\
\cmidrule{2-10}
& total & -0.189 & 6.6e-02 & -0.549 & \bfseries 6.8e-09 & 0.274 & \bfseries 6.8e-03 & -0.495 & \bfseries 2.9e-07 \\

\bottomrule
\end{tabularx}
    \end{subtable}
\end{table*}

\subsection*{\Harmful{} content spreads at night}
While the total number of posts per user is positively correlated with an increased ratio of \harmful{} content, heightened activity within a day is negatively correlated with spreading \harmful{} content at that time ($\rho = -.308$, \pvalue $=0.002$, \autoref{tab:diurnal_activity_corr}). This correlation is significant for all clusters except for evening types.

One's tendency to spread \harmful{} content shows temporal patterns beyond correlations with activity across the day.
We find particularly strong and regular distinctions between daytime and nighttime activity levels with respect to the spreading of \harmful{} content and the congruent content types (\autoref{tab:time_mannwhitneyu}). 
We analyse three different time periods: daytime and nighttime as defined by the clock,  by the presence of daylight, as well as by a user's regular and prolonged wakefulness times.

\begin{table*}
    \caption{Mann-Whitney U test comparing the distributions of content type ratios\textsuperscript{1} during different time periods: daytime and nighttime, times between and outside of sunrise and sunset, as well as regular and prolonged wakefulness times. \pvalue s that were found to be statistically significant at the $\alpha = .05$ level are highlighted in bold.
    \label{tab:time_mannwhitneyu}}
    \begin{threeparttable}[htb]
\begin{tabularx}{\textwidth}{ll*{6}{>{\raggedleft\arraybackslash}X}l}
\toprule
 &  & \multicolumn{2}{c}{6:30 am--6:45 pm\tnote{2,3}} & \multicolumn{2}{c}{sunrise--sunset\tnote{2,4}} & \multicolumn{2}{c}{waking--bedtime\tnote{2,5}} & \multirow[l]{2}{*}{lower\tnote{6}}\\
 \cmidrule(lr){3-4}\cmidrule(lr){5-6}\cmidrule(lr){7-8}
&  & $U$ & \pvalue{} & $U$ & \pvalue{} & $U$ & \pvalue{} &  \\

\midrule\multirow{4}{*}{\small \raggedright \Harmful{}}
 & infrequent & 647,659 & \bfseries 1.2e-05 & 654,543 & \bfseries 5.1e-05 & 553,578 & \bfseries 5.4e-03 & day \\
 & morning & 651,039 & \bfseries 1.4e-04 & 646,905 & \bfseries 6.8e-05 & 556,762 & \bfseries 3.2e-02 & day \\
 & intermediate & 539,718 & \bfseries 2.1e-06 & 544,018 & \bfseries 1.2e-05 & 348,336 & \bfseries 9.6e-07 & day \\
 & evening & 634,352 & \bfseries 3.2e-07 & 624,042 & \bfseries 1.7e-08 & 592,675 & 8.3e-01 & day \\
\midrule
\multirow{4}{*}{\small \raggedright Political}
 & infrequent & 575,508 & \bfseries 3.2e-09 & 587,627 & \bfseries 1.2e-07 & 514,388 & 5.5e-02 & day \\
 & morning & 485,640 & \bfseries 2.5e-13 & 505,402 & \bfseries 2.3e-10 & 371,236 & \bfseries 3.3e-16 & day \\
 & intermediate & 365,540 & \bfseries 3.1e-13 & 376,100 & \bfseries 2.3e-10 & 136,102 & \bfseries 9.3e-28 & day \\
 & evening & 516,744 & \bfseries 3.4e-17 & 527,524 & \bfseries 1.4e-14 & 501,261 & 4.3e-01 & day \\
\cline{1-2}
\multirow{4}{*}{\small \raggedright  Fake or hoax}
 & infrequent & 562,988 & 4.5e-01 & 571,088 & 2.5e-01 & 395,852 & \bfseries 5.0e-03 & day \\
 & morning & 472,360 & \bfseries 2.3e-04 & 487,556 & \bfseries 1.2e-02 & 333,468 & \bfseries 1.1e-05 & day \\
 & intermediate & 341,718 & \bfseries 7.5e-06 & 351,354 & \bfseries 3.9e-04 & 85,136 & \bfseries 6.8e-27 & day \\
 & evening & 543,092 & \bfseries 2.7e-04 & 540,036 & \bfseries 4.6e-05 & 475,132 & 1.1e-01 & day \\
\cline{1-2}
\multirow{4}{*}{\small \raggedright Conspiracy \& junk science}
 & infrequent & 535,523 & 7.4e-01 & 518,803 & 3.2e-01 & 397,920 & \bfseries 1.1e-04 & night \\
 & morning & 619,142 & \bfseries 1.6e-02 & 600,400 & 3.3e-01 & 514253 & \bfseries 1.0e-08 & night \\
 & intermediate & 428,288 & \bfseries 8.3e-03 & 408,088 & \bfseries 1.6e-04 & 149,368 & \bfseries 1.8e-08 & day \\
 & evening & 671,979 & \bfseries 2.6e-02 & 640,144 & 5.7e-01 & 493,674 & 4.3e-01 & night \\
\bottomrule
\end{tabularx}
        \begin{tablenotes}
            \item[1] defined in \autoref{eq:ratio}
            \item[2] We account for a safety margin of $s=1$ hour before and after each border value.
            \item[3] compares the distribution of ratios $r(t,c,f)$ for $t \in [\text{7:30 am} - \text{5:45 pm})$ (\enquote{day}) with those for $t \in [\text{7:45 pm} - \text{6:30 am})$ (\enquote{night}), considering the safety margin. The border values are sunrise and sunset times averages over the months, rounded to the closest quarter hour.
            \item[4] \label{sun} compares the distribution of ratios sunrise and sunset (\enquote{day}) with those between sunset and sunrise (\enquote{night}). The sunset and sunrise times are calculated geometrically using the average latitude and longitude for the users in our dataset for the first day of each month using Python's suntime library \url{https://github.com/SatAgro/suntime}. For users who only listed \enquote{Italy} as their location, the coordinates are approximated around the geographical centre of the peninsula. \label{footnote:suntime}
            \item[5] \label{waking} compares the distributions of ratios within 
            $[i(g(c,n),s), i(g(c,n),n-s))$ (\enquote{day})
            with those of the interval 
            $[i(g(c,n),n+s), i(g(c,n),-s))$ (\enquote{night}) 
            for $n=16$.
            $i(t,n)$ and $g(c,n)$ are defined in Equations \ref{eq:mod_day} and \ref{eq:heightened_activity}, respectively.
            \item[6] \label{lower} For each row, returns the distribution for which the corresponding \pvalue{} of a one-tailed Mann-Whitney U test was lower for all significant ($p < \alpha$) comparisons.
        \end{tablenotes}
    \end{threeparttable}
\end{table*}

\begin{figure*}
    \centering
    \caption{The ratio of \harmful{} content over time of day on the x-axis, and year and month on the y-axis. The darker a square, the higher the ratio of \harmful{} content. The hatched curves indicate the average sunrise and sunset times within a given month. The dotted lines represent the active times per cluster, and the times of day as defined by the clock. Missing values are presented in grey.\label{fig:heatmap_harmful}}
     \begin{subfigure}[t]{0.50\textwidth}
         \centering
         \caption{\Rcluster{}\label{fig:heatmaps_infr}}
         \includegraphics[width=\textwidth,keepaspectratio]{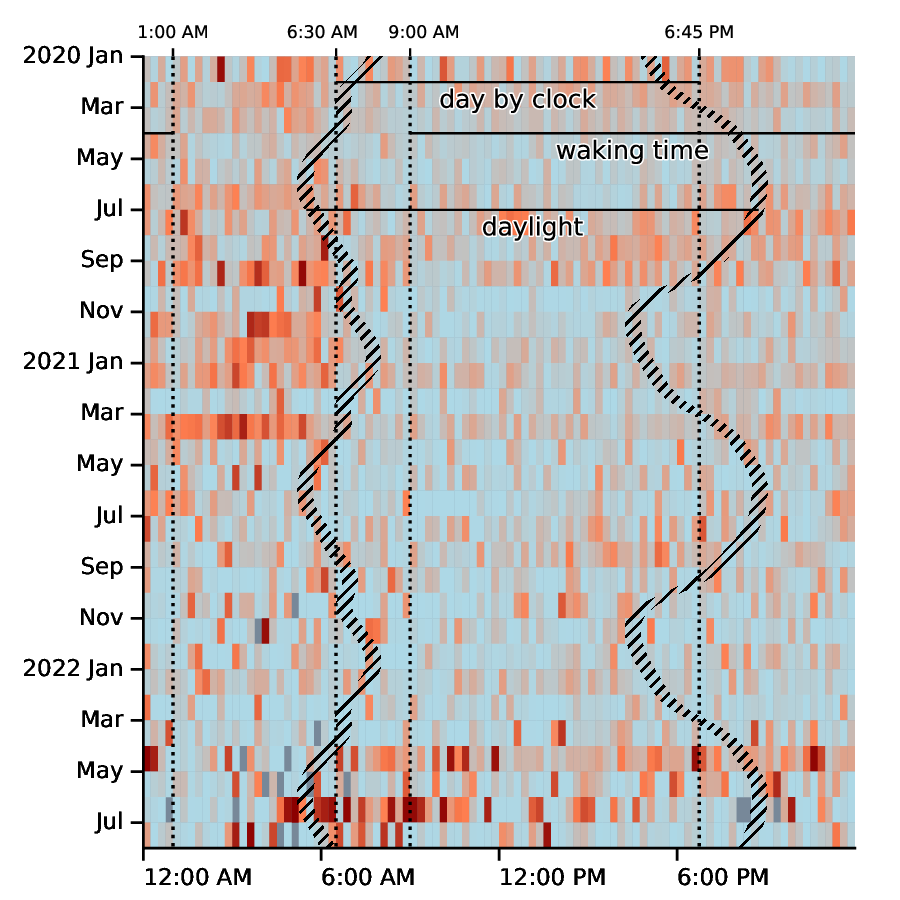}
     \end{subfigure}%
     \hfill%
     \begin{subfigure}[t]{0.50\textwidth}
         \centering
         \caption{Morning type\label{fig:heatmaps_morning}}
         \includegraphics[width=\textwidth,keepaspectratio] {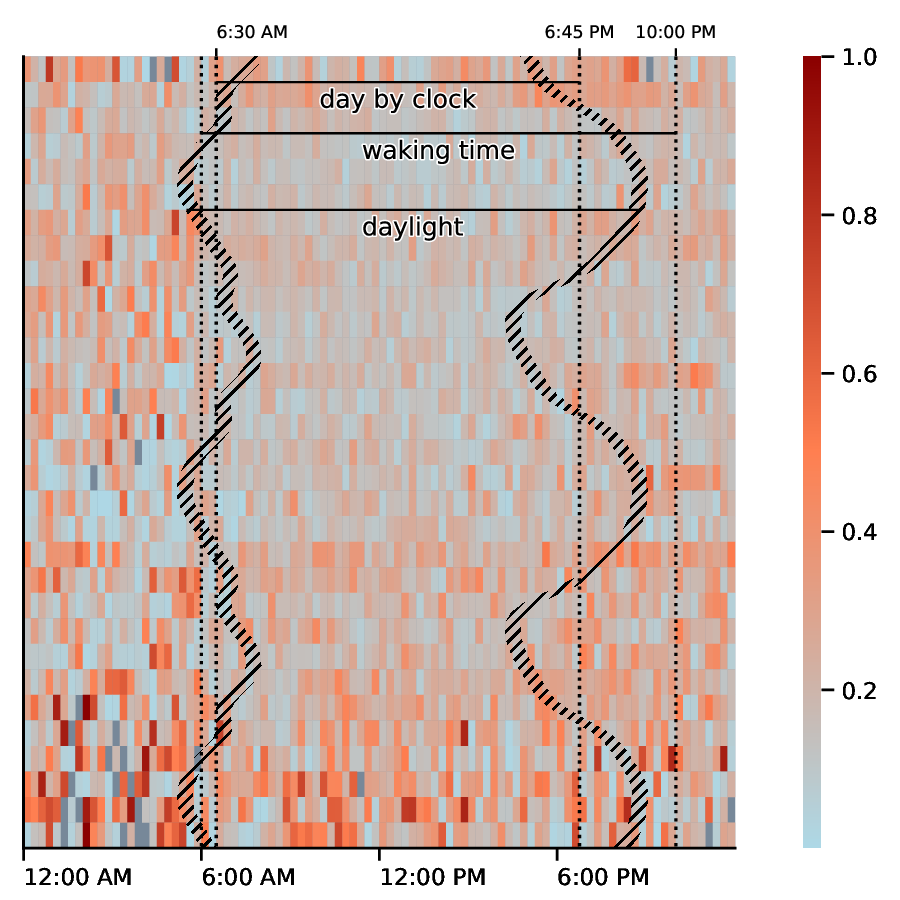}
     \end{subfigure}
     \vspace{0.5em}
     \begin{subfigure}[t]{0.50\textwidth}
         \centering
         \caption{\Scluster{}\label{fig:heatmaps_standard}}
         \includegraphics[width=\textwidth,keepaspectratio]{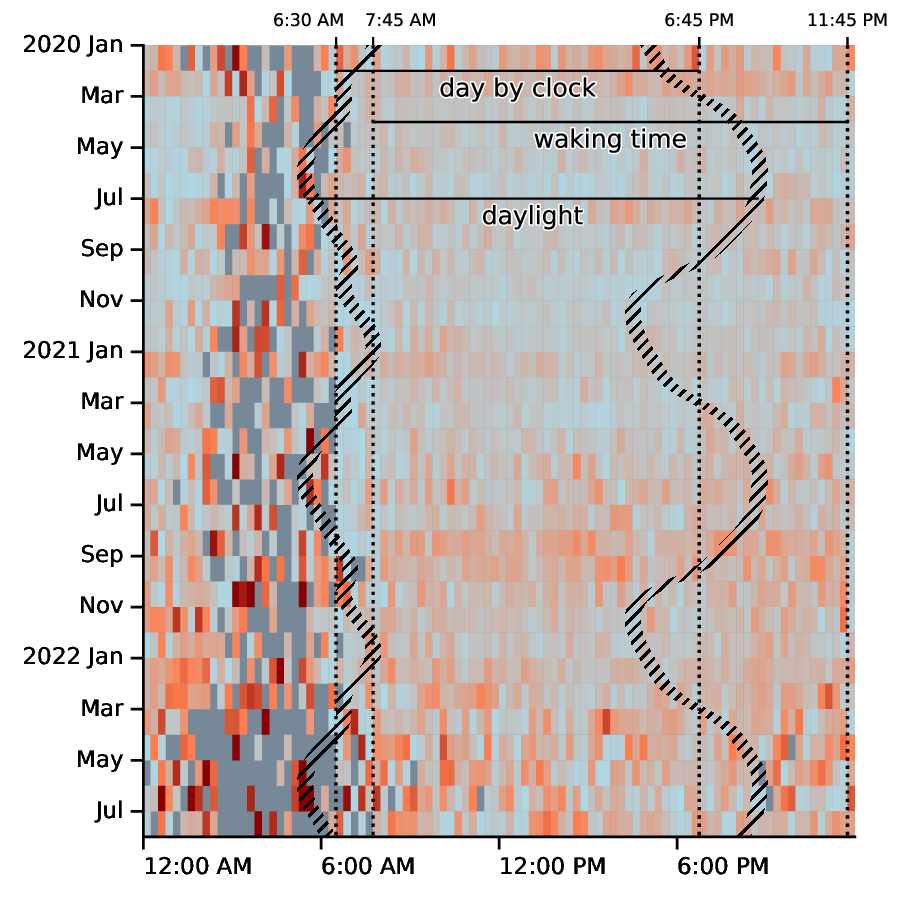}
     \end{subfigure}%
     \hfill%
     \begin{subfigure}[t]{0.50\textwidth}
         \centering
         \caption{Evening type\label{fig:heatmaps_evening}}
         \includegraphics[width=\textwidth,keepaspectratio]{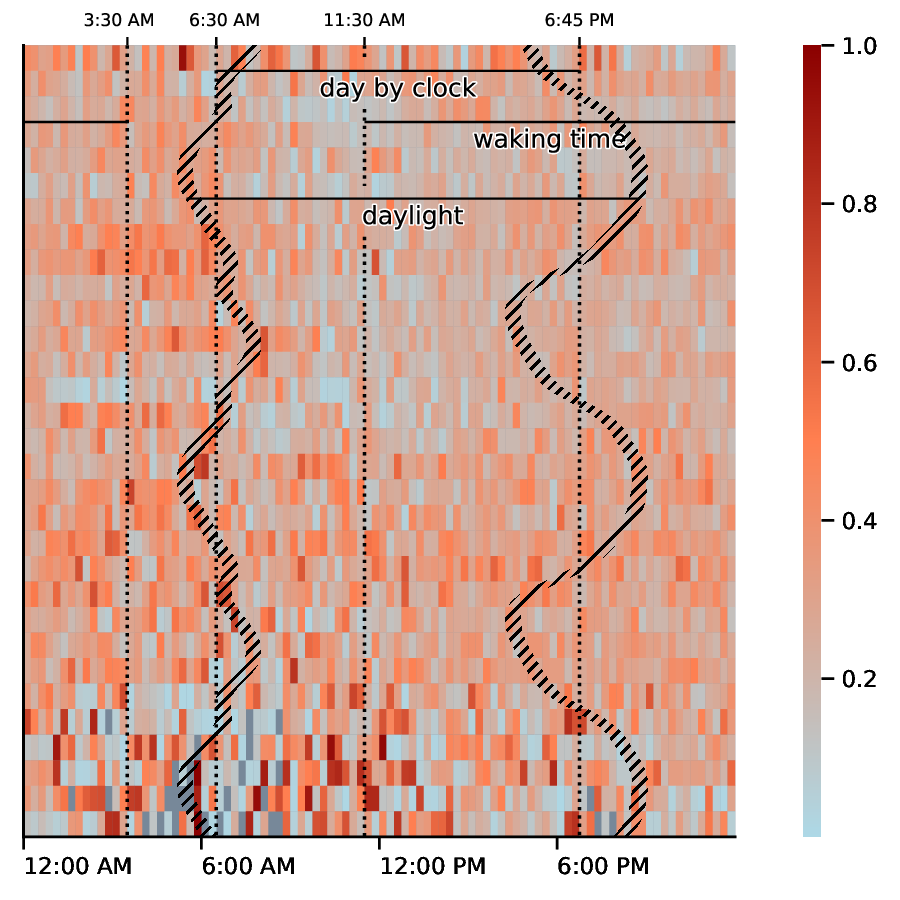}
     \end{subfigure}
\end{figure*}

\autoref{fig:heatmap_harmful} visually represents the comparisons between day and night periods for each cluster. The dotted vertical lines mark times of day as defined by the clock as well as regular waking times. The shaded areas represent the average sunrise and sunset times at the locations of the users in our dataset within Italy (which is helpfully vertical, with sunset and sunrise times differing by less than an hour at most in between any point on the map.) In our statistical analysis, we compare the time periods \enquote{within} these border with those \enquote{outside} them.

We find a statistically significant increase in the proportion of \harmful{} content shared between 6:45 pm and 6:30 AM for all clusters ($U >= 539,718$, \pvalue$<.001$). 
Similarly, more \harmful{} content is spread outside daylight hours for all clusters ($U >= 544,018$, \pvalue$<.001$). The increase during prolonged wakefulness is statistically significant for all clusters except \emph{evening types} ($U >= 348,336$, \pvalue$<.032$ for the other clusters).

\subsection*{\Harmful{} content adheres to clock time}

Time of day proves a stronger predictor than a user's activity throughout the day when looking at the continuum of \harmful{} content spread throughout the day.
For all clusters, ratios of \harmful{} content are highest in the early mornings in between 2:30 AM and 4:15 AM (\autoref{fig:fourier_ratios_clocktime}).
When aligned by waking time (\autoref{fig:fourier_ratios_waking}) the peak of \harmful{} content spreading falls across a wider time span, between 16:45 and 20:30 hours after waking (\supplref{tab:activity_ratio_stats}). 
Similarly, the distance of curves of \harmful{} content ratios robustly increased across several metrics as compared to an alignment by time of day (\supplref{tab:similarity_ratio_by_activity}). 
Therefore, the tendency to spread \harmful{} content seems to follow its own diurnal rhythm beyond the user's habitual use of Twitter.

\subsection*{Chronotypes prefer different content types}

We have so far analysed the binary categories of content that is \harmful{}, and content that is unlikely to be so. There are, however, also interesting observations within the individual content types. 

\begin{figure*}
    \centering
    \caption{
        Figure displaying, for each cluster
        the cumulative number of posts with known reliability classification throughout the day (coloured areas),
        the cumulative ratios of \harmful{} content types (red lines),
        the user's 8 least active hours (prolonged wakefulness, grey inner arc), and
        the times with the highest quantile of \harmful{} posts (red outer arcs).
        The axis scales are shared between panels.\label{fig:cluster_clockface}}
    \begin{subfigure}[b]{0.5\textwidth}
         \centering
         \caption{\Rcluster{}\label{fig:clockface_infr}}
         \includegraphics[width=\textwidth]{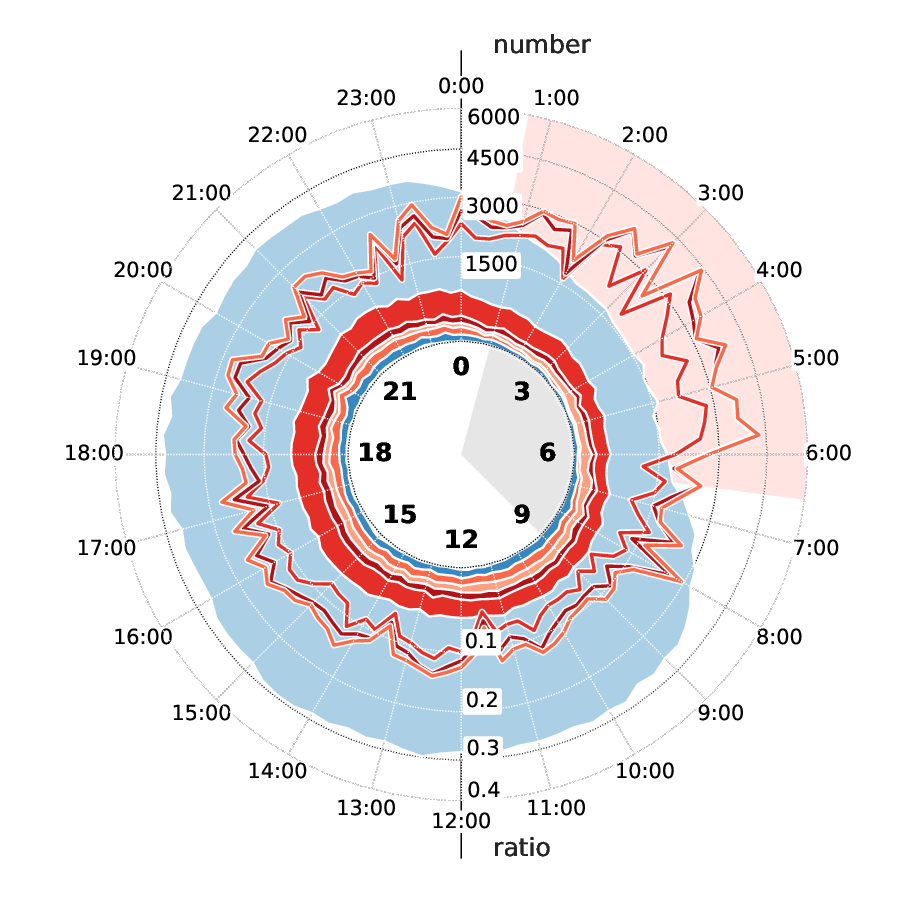}
     \end{subfigure}%
     \hfill%
     \begin{subfigure}[b]{0.5\textwidth}
         \centering
         \caption{Morning type\label{fig:clockface_morning}}
         \includegraphics[width=\textwidth]{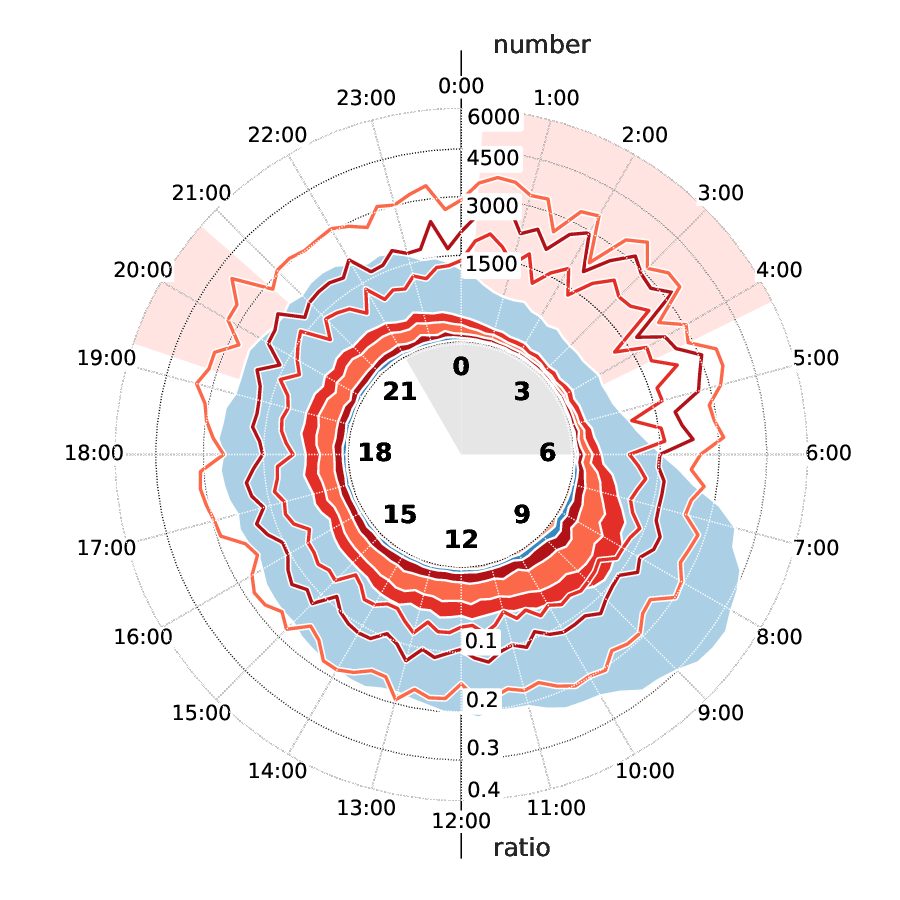}
     \end{subfigure}
     \vspace{0.5em}
     \begin{subfigure}[b]{0.5\textwidth}
         \centering
         \caption{\Scluster{}\label{fig:clockface_standard}}
         \includegraphics[width=\textwidth]{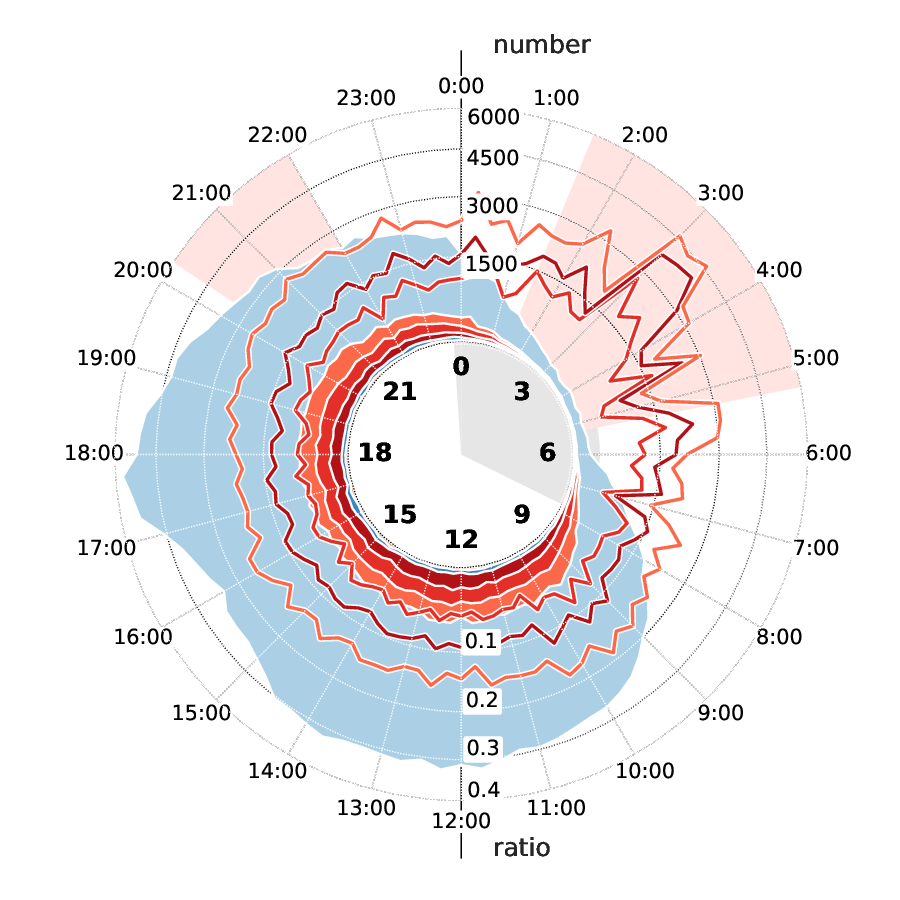}
     \end{subfigure}%
     \hfill%
     \begin{subfigure}[b]{0.5\textwidth}
         \centering
         \caption{Evening type\label{fig:clockface_evening}}
         \includegraphics[width=\textwidth]{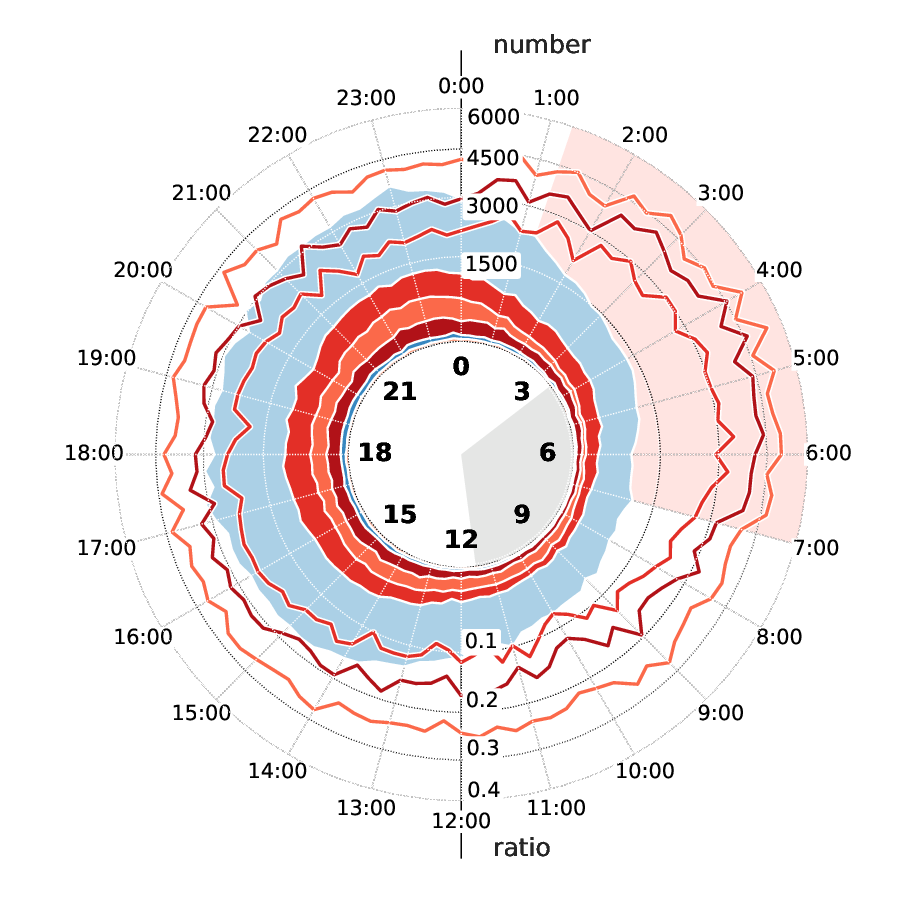}
     \end{subfigure}
     \begin{subfigure}[b]{\textwidth}
         \centering
         \includegraphics[width=\columnwidth]{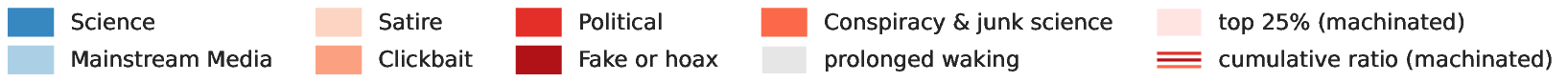}
     \end{subfigure}
\end{figure*}

The coloured areas of \autoref{fig:cluster_clockface} represent the activity of all user clusters and individual content types around a 24-hour clock. \emph{Morning} and \emph{evening types} show a particular tendency towards conspiracy theories and junk science, especially as compared to \emph{infrequent types}, who show the strongest inclination towards scientific content of all clusters. 
Only \emph{intermediate types} spread even more conspiracy and junk science than politically biased content (\supplref{tab:content_categories}). However, mainstream media reassuringly make up the vast majority of content spread by all clusters.

The red lines in \autoref{fig:cluster_clockface} represent the cumulative ratios of \harmful{} content types. For all clusters, the ratio of fake or hoax content increases noticeably during the nighttime when ratios of conspiracy and junk science are lowered. The two content types show opposite tendencies over the course of the day. The positive correlation of conspiracy theories and junk science with activity throughout the day is, however, only significant for \emph{infrequent} ($\rho = .263$, \pvalue $= .009$) and \emph{morning type} users ($\rho = .419$, \pvalue $>.001$, \supplref{tab:diurnal_activity_corr}).

\autoref{fig:cluster_clockface} also shows the times where one's tendency to spread \harmful{} content is in the top quartile ($Q_3$ in a 4-quantile) by red shading in the graph's background. The inner grey arcs represent the time of prolonged wakefulness for each cluster (see also \supplref{tab:dip_test_activity}). 
\emph{\Rcluster{}} experience the onset of increased spreading of \harmful{} content around their bedtime at 1:00 AM and only shortly before \emph{evening type} individuals. \emph{Evening types}, however, only enter prolonged wakefulness at 3:30 AM. For \emph{morning} and \emph{intermediate types}, the times of increased tendency to spread \harmful{} content is spread over different parts of the day, one within and one outside of prolonged wakefulness.

\subsection*{The impact of the lockdown}
As our dataset collects content related to the \covid{} pandemic, we must consider the impact of non-pharmaceutical interventions, such as home office or curfews, on daily rhythms as well as potential changes in the macroscopic informational landscape of Twitter \cite{Castaldo2021}. We specifically consider the time period of Italy's first lockdown from March 9\textsuperscript{th} to May 18\textsuperscript{th}, 2020.
During this time, as opposed to the entire span covered by the dataset, the ratio of \harmful{} content is lowered for all clusters by at least 2\%. 
However, all clusters tweeted more content featuring the \covid-related keywords in the dataset's search query during the lockdown period. 
\Harmful{} content was represented over-proportionally within this rise (\autoref{tab:stats_lockdown}, lockdown and number of posts of \harmful{} content are from different populations, $\chi^2=1344.17$, \pvalue $<.001$). 
The reduction of \harmful{} content ratios during lockdown can therefore be attributed to an increase in other content types, likely including a surge of informational coverage driven by mainstream and state media \cite{Gallotti2020}.

\begin{table*}
    \caption{Ratio of \harmful{} content during and outside of the lockdown period.}
    \label{tab:stats_lockdown}
    \begin{tabularx}{\textwidth}{>{\raggedright\arraybackslash}Xlrrrr}
\toprule
 &  & evening & infrequent & intermediate & morning \\
\midrule
\multirow[l]{3}{16em}{\harmful{} posts per day and user} & no lockdown  & 0.003 & 0.068 & 0.077 & 0.064 \\
 & lockdown & 0.007 & 0.128 & 0.121 & 0.105 \\
 & change & 0.004 & 0.060 & 0.044 & 0.041 \\
\cline{1-2}
\multirow[l]{3}{16em}{posts per day and user} & no lockdown & 0.000 & 0.017 & 0.013 & 0.021 \\
 & lockdown & 0.001 & 0.024 & 0.014 & 0.031 \\
 & change & 0.000 & 0.008 & 0.001 & 0.010 \\
\cline{1-2}
\multirow[l]{3}{16em}{ratio of \harmful{} content} & no lockdown & 0.149 & 0.194 & 0.157 & 0.276 \\
 & lockdown & 0.128 & 0.156 & 0.113 & 0.244 \\
 & change & -0.021 & -0.037 & -0.044 & -0.032 \\
\bottomrule
\end{tabularx}

\end{table*}

\section*{Discussion}

Propaganda campaigns and targeted manipulation continue to endanger our cognitive autonomy and unhampered opinion formation \cite{Lin2019}.
With Large Language Models tapping into an unrivalled potential to scale the generation and deployment of mis- and disinformation, the factors impacting our susceptibility and reaction thereto are at risk of and may well already be subject to exploitation.
A deeper scientific understanding of user response to \harmful{} content can, however, also aid in the prevention of an unwitting contribution to such campaigns. 

Specifically, we extrapolate two main takeaways from our study:
First, user activity on social media throughout the day can be mapped to chronotypical archetypes on the morningness-eveningness continuum. We find these activity patterns to be a predictor of one's propensity to spread \harmful{} content and the constituent content types. \emph{Evening types} have the significantly highest inclination towards \harmful{} content, \emph{\rcluster{}} the lowest.
Secondly, the spread of \harmful{} content is linked to time of day more so than to activity patterns by user type, reaching a peak between 2:30 AM and 4:15 AM.

These lessons have implications for (a) our understanding of user responses to \harmful{} information in relation to user activity and time of day, and (b) the design of interventions to prevent the spread of mis- and disinformation on social media.

Generally, our findings are in line with previous literature detailing the link between cyclical behavioural patterns and Twitter use~\cite{Kates2021, Dzogang2018, Mayor2021, Dzogang2017} as well as with findings associating sunlight with cognitive function (and by extension critical thinking) \cite{Kent2009} and with activity on Twitter \cite{Leypunskiy2018, Gleasure2020}. 
There is, however, a remarkable distinction between the diurnal activity curves and the curves of ratio of \harmful{} content spread by the clusters. The former exhibit a significant similarity in peak activity  times (and in the time of activity trough) when aligned by waking time. The latter, in contrast, shows a higher closeness of peak ratios under clock time rather than considering time after waking, occurring between 2:30 AM and 4:15 AM.
This suggests that the likelihood of spreading \harmful{} content is more than a function of increasing tiredness, though indeed, prolonged wakefulness is known to impact cognitive performance~\cite{Alhola2007}. The time interval in question --- the hour between 3 AM and 4 AM is fittingly known in common parlance as the witching hour \cite{Luke2014} --- coincides with the approximate peak in melatonin (3 AM \cite{Stehle2011}) and lymphocyte levels (4 AM \cite{Suzuki2003}), as well as the troughs of epinephrine and norepinephrine (3:30 and 2:30 AM, respectively \cite{Linsell1985}). There may, therefore, exist a direct physiological link between the time of day and susceptibility to mis- and disinformation. 

Our research may inform the timing of interventions against mis- and disinformation, and concentrate efforts on limited time frames.
Continuously deploying interventions may be more costly for the implementer and may overload the user's attentional capacity and patience. Shorter exposition may be more resource-effective and less intrusive. 
Impactful times may include the peak activity times of those users most susceptible to \harmful{} content (such as around 10 pm to target individuals with an evening preference) or times when users are most likely to spread \harmful{} content (such as around 3 AM).
The potential of our findings to inform the design of protective measures is all the more relevant in light of the rising trend in cyber operations and information warfare \cite{Lin2019,Mazarr2019}.

More specifically, in the context of \covid{}, the non-pharmaceutical interventions imposed by many countries, such as lockdowns, curfews and home office, have disrupted many peoples' daily rhythms, plausibly giving rise to interaction effects between circadian mismatch and the course of the pandemic \cite{Romigi2022}. 
We do not find an increased spread of the ratio of \harmful{} content shared during lockdown. However, the outcome of continued measures such as home-office or curfews may well have aided the related spread of conspiracy theories \cite{Dreyfuss2017, Roozenbeek2020}.
Future policy interventions should therefore consider their possible impact on human circadian activity to limit the risk of concomitant  increases in mis- and disinformation \cite{Gallotti2020}.

While a social media study allows the analysis of social dynamics at an unprecedented scale, it also comes with a set of limitations. In particular, using a dataset collected entirely from Twitter biases the reference population towards being more highly educated, working age, and male. The dataset, alongside its limitations, is discussed in detail in Gallotti \textit{et al.} (2020)~\cite{Gallotti2020}. 
In terms of analysis, we use a set of proxy metrics: the ratio of \harmful{} content (as a proxy for susceptibility to mis- and disinformation), activity patterns on Twitter (as a proxy for the user's chronotype), and average times of sunset and sunrise (as a proxy for sunlight exposure). These are computationally viable options allowing the large-scale analysis of behavioural phenomena but cannot measure the phenomena directly.

Several questions and challenges remain unanswered by this study. 
Causality is yet to be established for the impact of time of day, chronotype, and non-pharmaceutical interventions against \covid{} on one's susceptibility to mis- and disinformation. Controlled behavioural experiments, in particular, would allow us to consider more direct measures of the proxies we defined above. 
Further challenges include an extension and comparison across countries, languages, platforms, and representative user groups.

On a larger scale, we hope for further research into how knowledge of the diurnal patterns of our reaction to mis- and disinformation can effectively be leveraged and integrated into the design of interventions against large-scale manipulation. Temporality, along with other factors impacting out susceptibility to mis- and disinformation, are likely already modeled in the latent space of deep learning systems. An analytic understanding can aid us in maintaining integrity of mind and autonomy of thought.

\section*{Methods \label{sec:methods}}
\subsection*{Data} \label{methods:data}

We consider a Twitter dataset~\cite{Gallotti2020} collected through the Twitter Filter API based on a set of hashtags and keywords surrounding the Covid-19 pandemic, specifically \emph{coronavirus, ncov, \#Wuhan, covid19, covid-19, sarscov2, covid}. 
Analysis was limited to the time span of January 22, 2020, when more than 6000 cases were reported in China, up to August 1st 2022. Twitter restrictions limit collection to no more than 4.5 million messages per day, on average. 
9,128 tweets collected between January and February 2021 were not associated with a tweet type on collection and were excluded from analysis. 
After removal of duplicates and posts by users identified as bots, our body of analysis encompassed 18,148,913 tweets, retweets or replies, of which 1,001,045 are assigned a known reliability.

\subsection* {Source reliability mapping}
Tweets were assigned a source reliability rating by the dataset authors \cite{Gallotti2020} based on manually checked web domains from multiple public databases, including journalistic and scientific sources \cite{Zimdars2016, Silverman2017, FakeNewsWatch, Politifact2017, Bufale2018, Starbird2018, Fletcher2018, Grinberg2019, MediaBiasFactCheck}. From these sources, the authors created a database of 3892 domains after cleaning and processing. Tweets containing a link are compared to domains in the database and classified according to domain reliability. The categories were adapted to fit the project focus and are detailed in \supplref{tab:content_categories}.

\subsection* {Geographic and time zone mapping}
Geocoding and geodata cleaning was conducted by the dataset authors \cite{Gallotti2020} based on the user's self-declared location field \emph{ArcGIS API}. 
Mapping errors (based, for example, on non-toponymos entries or website URLs) entries were removed by isolating single locations associated with many different unique location strings and data restricted to country-based granularity.
Within this study, we use exclusively the data found to originate from Italy. By extension, we ported the time zone of content returned by the Twitter API to Central European Summer or Winter Time, respectively. 

\subsection*{Clustering} \label{methods:clustering}

Let $T=\{[t, t+\frac{1}{4})\ |\ 4t \in \mathbb{N} \wedge \ 0 \leq t < 24 \}$ be the set of 15 minute intervals within a day given in hours, $F$ the set of content types and $I$ the set of users authoring content.
We will subsequently use $t$ to refer to one such interval $[t, t+\frac{1}{4}) \in T$ for simplicity.
Let then $\{ P_{(t,i,f)} \}_{(t, i, f) \in T \times I \times F}$ be the set of posts of content type $f \in F$ authored during interval $t \in T$ by user $i \in I$, indexed by a surjective function from $T \times I \times F$ onto $P$.

We cluster users based on their average posting activity levels during an interval $t \in T$:
\begin{equation} \label{eq:activity_a_ti}
    a(t,i)=\frac{\sum_{f \in F} \abs{P_{(i,t,f)}}}{\sum_{t \in T}\sum_{f \in F}\abs{P_{(i,t,f)}}}
\end{equation}
The activity levels were smoothed using a rolling average over a 90 minute Gaussian window, looping the values around midnight.

Six cluster performance indicators (specifically, Elbow \cite{Thorndike1953}, Context-Independent Optimality \cite{Gurrutxaga2010}, Cali\~{n}ski-Harabasz \cite{Calinski1974}, Davies-Boulin \cite{Davies1979}, generalised Dunn \cite{Dunn1973} and Silhouette \cite{Rousseeuw1987}) informed our choice of cluster method and number of clusters. We applied agglomerative hierarchical clustering with Ward's Minimum Variance method \cite{Balcan2014}.
An initial analysis revealed the presence of six distinct clusters with unique patterns of posting activity. We verified to receive similar clusters when considering posts when considering only unverified users (\ref{sec:suppl:unverified}). One of these clusters ($69$ users) showed suspicious bot-like activity with high levels of activity narrowly distributed around 10 AM, and posting almost exclusively content including links that are anonymised and often temporary for higher obscurity. We filtered out those users of this cluster who were classified as bots by Botometer\cite{Yang2022} (25 users) and subsequently repeated the clustering procedure. This resulted in three distinct clusters with unique patterns of posting activity (\emph{morning, intermediate} and \emph{evening type} posters). Users with low post rates ($<240$ posts) are separated into a fourth cluster (\emph{infrequent type posters}).

Inter- and intra-cluster distances are detailed in~\supplref{tab:similarity_ratio_by_activity}, 
general information about the clusters is given in~\supplref{tab:cluster_stats}.

\subsection*{Diurnal activity} \label{methods:diurnal_activity}

Let $C$ be the set of all clusters where $c \in C$ is a subset of $I$. Function 
\begin{equation} \label{eq:diurnal_activity_coarse}
    a_c(t)=\frac{\sum_{i \in C}\sum_{f \in F} \abs{P_{(i,t,f)}}}{\sum_{t \in T}\sum_{i \in C}\sum_{f \in F}\abs{P_{(i,t,f)}}}
\end{equation}
calculates the activity levels during an interval $t$ by cluster $c$. To denoise and compare the cluster activity curves, we transform them from the time domain into the frequency domain using the discrete Fourier transform:

\begin{equation}
    X^c_k = \sum^{N-1}_{n=0} a_{c,n} e^{-\frac{i 2 \pi}{N} kn}
    \qquad 
    k \in [0, N-1]
    \label{eq:dft}
\end{equation}

where $a_{c,n}=a_c(t_k)$ and $t_k = k\Delta$.
\autoref{eq:dft} yields a sequence of complex numbers $\{X_k^c\} = X^c_0, X^c_1, ..., X^c_{N-1}$ which describe amplitude and phase of sinusoidal functions. On summation, the sequence produces the original discrete signal. In particular, the $k^{th}$ Fourier coefficient provides information about the sinusoid that has $k$ cycles over the given number of samples. 

We then identified the coefficients with the greatest amplitude. Let $\{A^c \} = \{A^c_1, A^c_2, ..., A^c_{N-1}\}$ be the set of all amplitudes of the constituent sinusoidal functions for frequencies $0, 1, ..., N$, and let $\{ A^{(c,m)} \} \subset \{A^c \}$ be the set of $m$ largest amplitudes.

The signal is then recombined as follows to contain only the harmonics with $m$ greatest amplitudes:

\begin{align} \
    h^c(n, t) &=A^c_n \cos{\frac{2\pi}{P^c}nt - \varphi_n^c} \label{eq:harmonic} \\
    S^{c,m}_N(t) &\approx \frac{A^c_0}{2} + \sum^{N}_{n=1} \left \{
    \begin{aligned}
        & h^c(n, t) && \text{if} \quad A^c_n \in \{ A^{(c,m)} \} \\
        & 0 && \text{otherwise}
    \end{aligned} \right. \label{eq:recomposed_signal}
\end{align}

where $h(n, t)$ describes the $n^{th}$ harmonic of the Fourier series. $P^c$ is the period of function $a(t,c)$, $A^c_n$, $\varphi^c_n$ and $\frac{n}{P^c}$ are amplitude, phase and frequency of harmonic $h^c(n, t)$ respectively, 
 and $S^{(c,m)}_N(t)$ approximates the recomposed signal at time point $t$. 
 
We used the value for $m$ where the change in distance to the next larger value grew smaller for each cluster. It two values are supported by an equal number of indicators, we chose the smaller one. 
Let $\{ U \}$ be a set of 6 distance metrics, specifically Partial Curve Mapping \cite{Witowski2012}, the area method \cite{Jekel2019}, discrete Frechet distance \cite{Frechet1906}, curve length \cite{AndradeCampos2012}, Dynamic Time Warping \cite{Berndt1994} as well as mean absolute error and mean squared error.
Let then $ \{ D_{u}^m \} = { \sum_{t \in T} u(S^{(c,m)}_N(t), a^c(t)) } $ describe the distances between the  original signal and the reconstruction (see \autoref{eq:activity_a_ti} and \autoref{eq:recomposed_signal}, respectively) for a given value of $m$ and a distance metric $u \in U$.
For a cluster $c$, we find the value of $m$ as:
\begin{align}
    m^c = \min \{ \text{mode} \{ \underset{m \in [1,4]} {arg min} \ (D_u^{(c,m)}, D_u^{(c,m-1)} ) \} \} \label{eq:finding_m}
\end{align}

where $\underset{m \in M}{arg min} \ h(m) = \{ m \ | \ h(x) \geq h(m) \ \forall x \in M \}$ returns the set of points $m$ for which a function $h(m)$ returns the function's smallest value, if it exists. The $\text{mode}$ operation returns the set of most common elements, and $\min$ finds the minimum element of a set.
We accordingly used$m=3$ for all clusters.

This leaves us with set $ \{ A_{(t,c)} \}_{(t,c) \in T x C}$ of smoothed diurnal cluster activity. Details on the maxima and minima are found in \supplref{tab:activity_ratio_stats}.

\subsection*{Periods of heightened activity and prolonged wakefulness} \label{methods:heightened_activity}

To find the periods of heightened activity, let 

\begin{equation} \label{eq:mod_day}
    i(t,n) = (t+n)(\bmod \ 24)
\end{equation}
return the time of day $n$ hours past $t$ where $\bmod$ refers to the modulo operator. Then, let

\begin{equation} \label{eq:t_within_n}
j(t,s,n) = 
\begin{cases}
    t < s \wedge s < i(t,n) & \text{if } t < i(t,n)\\
    s > t \vee s < i(t,n)   & \text{otherwise}
\end{cases}
\end{equation}

indicate whether a time point $s$ occurs within $n$ hours past $t$.
Then, the onset of heightened activity for cluster $c$ and for $n=16$ is found by:

\begin{equation} \label{eq:heightened_activity}
    g(c, n) = \underset{t \in T}{arg max} \ \sum_{s \in T \wedge j(s,t,n)} A_{(s,c)}
\end{equation}

Analogously to the $arg min $ operation, the set of points $t$ for which a function $h(t)$ returns the function's largest value, if it exists, is found as:
\begin{equation} 
\underset{t \in T}{arg max} \ h(t) = \{ t \ | \ h(x) \leq h(t) \ \forall x \in T \}
\end{equation}
 
The end of the period of heightened activity is then $i(g(c,n),n)$. 
\supplref{tab:dip_test_activity} lists these times for each cluster. We refer to the period after the end but before the onset of heightened activity as \emph{prolonged wakefulness}.

\subsection*{Weighted ratios of content types}\label{methods:weights}

Posts are weighted inversely to the total posts per authoring user, with the weight of a given post by user $i$ defined as 
\begin{equation} \label{eq:post_weight}
    w(i) = \frac{1}{\sum_{t \in T}\sum_{f \in F}\abs{P_{(t,i,f)}}}.
\end{equation}

We calculate the ratio of a given content type without including the category \enquote{Other}, which is not easily classifiable, makes up the vast majority of content in our dataset, and could possibly obstruct patterns in the data. 

Let therefore $F^K$ be the subset of $F$ without \enquote{Other}. The ratio for content type $f \in F^K$, cluster $c$ and 15 minute time interval within a day $t$ is calculated as 

\begin{equation} \label{eq:ratio}
    r(t, c, f) = \frac{\sum_{i \in c} \abs{P_{(t,i,f)}} \ w(i)} %
    {\sum_{g \in {F^K}} \sum_{i \in c} \abs{P_{(t,i,g)}} \ w(i)}.
\end{equation}

The ratio of \harmful{} content is then 

\begin{equation} \label{eq:ratio_harmful}
    r^H(t, c) = \frac{\sum_{f \in F^H} \sum_{i \in c} \abs{P_{(t,i,f)}} \ w(i)} %
    {\sum_{g \in {F^K}} \sum_{i \in c} \abs{P_{(t,i,g)}} \ w(i)}.
\end{equation}

where $F^H$ is the set of \harmful{} content types, consisting of conspiracy or junk science, fake or hoax news, and politically biased news.

In this way, each user carries the same weight across the dataset.

We applied the process described by Equations~\ref{eq:activity_a_ti} to~\ref{eq:finding_m} also to the diurnal pattern of ratios of \harmful{} content. On these curves, the values of $m$ for \autoref{eq:recomposed_signal} preceding the lowest change in distance metrics were $m=3$ for \emph{\scluster{}} users, and $m=2$ for all other types. We refer to the set of smoothed diurnal ratios of \harmful{} content as $ \{ R_{(t,c)} \}_{(t,c) \in T x C}$.

We consider a time span $t$ to reflect an increased susceptibility to spreading \harmful{} content for a given cluster if the smoothed ratio $R_{(t,c)}$ is greater than the third quartile. So $t$ is a time of increased susceptibility for cluster $c$ if $ Pr[\{ R_{(s,c)} | s \in T \} < R_{(t,c)}] \leq 3/4 $, where $Pr$ refers to the probability of an occurrence.

\subsection*{Statistics}\label{methods:statistics}

$\chi^2$-test was used for comparison of nominal variables, i.e. the relationship in between times of lockdown and \harmful{} content and in between content type and cluster affiliation. We  used the Dip Test of Unimodality \cite{Hartigan1985} to test unimodality of distributions of diurnal activity for each cluster. Unimodality could be rejected for all clusters both for the smoothed diurnal activity curves of set $ \{ A_{(t,c)} \}_{(t,c) \in T x C}$ and for the raw activity aggregations over the day described by \autoref{eq:diurnal_activity_coarse}. See \supplref{tab:dip_test_activity} for the Dip statistic and \pvalue s per cluster.

While we assume a monotonic relationship between the number of posts per user and the ratio of \harmful{} content, we do not assume a linear one. Therefore, we use Spearman's $\rho$ to describe correlation between these variables (\autoref{tab:total_posts_corr}). The same is true for correlation of user activity throughout the day with ratio of \harmful{} content throughout the day. \autoref{tab:diurnal_activity_corr} shows the correlation coefficient and \pvalue{} for the raw activity aggregations over the day and for the smoothed activity curves.

Neither diurnal activity nor diurnal ratio of \harmful{} content types are normally distributed (Shapiro–Wilk $W=.875$, \pvalue{}$>.001$ and $W=.886$, \pvalue{}$>.001$, respectively). Therefore, we used the nonparametric Mann-Whitney $U$ test to assess the difference in distributions of ratios of \harmful{} content throughout the day by cluster (\autoref{tab:mannwhitneyu_contentypes}) and between day and nighttimes (\autoref{tab:time_mannwhitneyu}).

\bibliography{arxiv.bib}

\section*{Acknowledgements}

The authors would like to thank the HumanE-AI-Net project, which has received funds from the European Union’s Horizon 2020 research and innovation programme under grant agreement 952026. RG acknowledges the financial support received from the European Union's Horizon Europe research and innovation program under grant agreement 101070190.
We thank Dino Carpentras, Dirk Helbing, Giulia Dalle Sasse and Manlio De Domenico for the valuable discussions and insights.

\section*{Author contributions statement}

All authors conceived and designed the experiments and wrote and reviewed the manuscript. E.S. performed the experiments, E.S. and C.I.H. analysed the results, R.G. contributed materials and analysis tools. 

\section*{Additional information}


The authors declare no competing interests.
\clearpage
\appendix
\appendixpage

\renewcommand{\thefigure}{S\arabic{figure}}
\renewcommand{\thetable}{S\arabic{table}}
\setcounter{figure}{0}
\setcounter{table}{0}
\pagenumbering{roman}
\setcounter{page}{1}

\captionsetup[table]{name=Supplementary Table}
\captionsetup[figure]{name=Supplementary Fig.}

\begin{sidewaystable}
    \centering
    \caption{Classification of social media content adapted from Gallotti \textit{et al.} (2020)\cite{Gallotti2020}. The term \enquote{\harmful{}} refers to those categories of  concern to democratic opinion formation. Each content category is listed alongside general statistics and ratio of posts per cluster. Ratios are negatively weighted by the author's total number of posts. $\sum_{t \in T} r(t,c,f)$ for a cluster $c$ and a content type $f$, as defined in \autoref{eq:ratio}. Ratios of \harmful{} content are notably elevated for \emph{evening types} compared to all other clusters.}
    \label{tab:content_categories}
    
\begin{tabularx}{\linewidth}{>{\raggedright}p{6em}>{\raggedright}X *{7}{r}}

\toprule
 Category
 & Characteristics
 & \multirow{2}{4em}{\centering total posts} 
 & \multirow{2}{5em}{\centering mean posts per author} 
 & \multirow{2}{6em}{\centering median posts per author} 
 & \multicolumn{4}{c}{ratio} \\
 \cmidrule{6-9}
&&&&& infrequent & morning & intermediate & evening \\
\midrule

    Science & subject to a rigorous validation process by scientific methods & 18,831 & 2,261 & 484 & 0.028 & 0.021 & 0.017 & 0.020 \\
    Mainstream media & subject to fact checking and media accountability & 757,467 & 2,683 & 672 & 0.743 & 0.780 & 0.821 & 0.695 \\
    Satire & distorts or misrepresents information for entertainment value, usually is easily identified & 4,301 & 734 & 170 & 0.008 & 0.004 & 0.005 & 0.004 \\
    Clickbait &  attempts to pass fabricated to misrepresented information as facts & 12,197 & 735 & 39 & 0.076 & 0.008 & 0.005 & 0.008 \\
    Other & general-purpose category collecting content which is not easily classifiable, includes links that are anonymised and often temporary for higher obscurity (originally \enquote{Shadow}), or does not contain links at all & 17,147,868 & 1,775 & 364 & - & - & - & -\\

\midrule 
     Political & aims to build a consensus on a polarised position by omission, manipulation and distortion of information & 98,700 & 2,755 & 721 & 0.107 & 0.078 & 0.057 & 0.146 \\
     Fake or hoax & entirely fabricated or manipulated content that aims to be perceived as realistic and reliable & 43,888 & 2,601 & 1,143  & 0.027 & 0.040 & 0.037 & 0.059 \\
     Conspiracy \& junk science & strongly ideological, inflammatory content alternative or oppositional to tested and accountable knowledge and information with the intent of building echo chambers & 65,661 & 4,275 & 1,679 & 0.012 & 0.070 & 0.058 & 0.067\\
     
    \cmidrule{2-9}
    \itshape\Harmful{} & \itshape composite category of politically biased information, fake or hoax news, and conspiracy and junk science & \itshape208,249 &
    \itshape 3,202 & \itshape 1,110& \itshape 0.146 & \itshape 0.188 & \itshape 0.152 & \itshape 0.272 \\ 
    \cline{3-9}
    \bottomrule
\end{tabularx}
\end{sidewaystable}

\begin{table}
    \centering
    \caption{Times of maximum and minimum activity as well as ratios of \harmful{} content per cluster sorted by extremity, (i.e. the first row per cluster shows the largest maximum and smallest minimum).}
    \label{tab:activity_ratio_stats}

\begin{tabularx}{\linewidth}{
>{\raggedright\arraybackslash}p{1.5em} 
>{\raggedright\arraybackslash}p{5em} 
l
Xrr>{\raggedleft\arraybackslash}Xr}
\toprule
& & \multicolumn{3}{c}{max} & \multicolumn{3}{c}{min} \\
  \cmidrule(lr){3-5}\cmidrule(lr){6-8}
& & clock time & hours past waking & activity/ratio & clock time & hours past waking & activity/ratio \\
\midrule
\multirow{8}{*}{\rotatebox[origin=c]{90}{\small activity}}
& \multirow[l]{2}{*}{infrequent} & 19.00 & 10.00 & 0.013 & 5.75 & 20.75 & 0.007 \\
& & 14.25 & 5.25 & 0.012 & 16.00 & 7.00 & 0.012 \\
\cline{2-2}
& \multirow[l]{2}{*}{morning} & 9.25 & 3.25 & 0.020 & 3.25 & 21.25 & 0.001 \\
& & 16.75 & 10.75 & 0.012 & 15.00 & 9.00 & 0.011 \\
\cline{2-2}
& \multirow[l]{2}{*}{intermediate} & 12.00 & 4.25 & 0.018 & 4.75 & 21.00 & 0.000 \\
& & 18.25 & 10.50 & 0.017 & 15.50 & 7.75 & 0.016 \\
\cline{2-2}
& \multirow[l]{2}{*}{evening} & 22.25 & 10.75 & 0.017 & 9.00 & 21.50 & 0.005 \\
& & 16.00 & 4.50 & 0.013 & 17.25 & 5.75 & 0.013 \\

\midrule
\multirow{11}{*}{\rotatebox[origin=c]{90}{\small ratio}}
& \multirow[l]{2}{*}{infrequent} & 3.75 & 18.75 & 0.226 & 11.00 & 2.00 & 0.096 \\
& & 16.75 & 7.75 & 0.154 & 21.00 & 12.00 & 0.133 \\
\cline{2-2}
& \multirow[l]{3}{*}{morning} & 2.50 & 20.50 & 0.222 & 14.50 & 8.50 & 0.167 \\
& & 20.00 & 14.00 & 0.211 & 8.00 & 2.00 & 0.178 \\
& & 10.50 & 4.50 & 0.184 & 22.50 & 16.50 & 0.204 \\
\cline{2-2}
& \multirow[l]{4}{*}{intermediate} & 3.50 & 19.75 & 0.206 & 13.50 & 5.75 & 0.130 \\
& & 21.25 & 13.50 & 0.180 & 7.00 & 23.25 & 0.140 \\
& & 9.75 & 2.00 & 0.169 & 17.25 & 9.50 & 0.141 \\
& & 16.25 & 8.50 & 0.142 & - & - & - \\
\cline{2-2}
& \multirow[l]{2}{*}{evening} & 4.25 & 16.75 & 0.316 & 11.50 & 0.00 & 0.234 \\
& & 17.50 & 6.00 & 0.276 & 21.75 & 10.25 & 0.262 \\
\bottomrule
\end{tabularx}
\end{table}

\begin{table}
    \centering
    \caption{Dip-test of modality of user activity curves (panels \enquote{coarse} and \enquote{smooth}) as well as times of onset and end of heightened activity per cluster. Significant results (\pvalue < .005) are given in bold font.}
    \label{tab:dip_test_activity}
    \begin{tabularx}{\textwidth}{l*{7}{>{\raggedleft\arraybackslash}X}}
\toprule
 & \multicolumn{2}{c}{coarse} & \multicolumn{2}{c}{smooth} & \multicolumn{2}{c}{heightened activity} \\
 \cmidrule(lr){2-3}\cmidrule(lr){4-5}\cmidrule(lr){6-7}
 
& dip statistic & \pvalue{} & dip statistic & \pvalue{} & onset & end \\
\midrule
infrequent & 0.052 & \bfseries 0.039 & 0.059 & \bfseries 0.013 & 9:00 AM & 1:00 AM \\
morning & 0.028 & 0.908 & 0.063 & \bfseries 0.007 & 6:00 AM & 10:00 PM \\
intermediate & 0.060 & \bfseries 0.009 & 0.082 & \bfseries 0.001 & 7:45 AM & 11:45 PM \\
evening & 0.068 & \bfseries 0.001 & 0.085 & \bfseries 0.001 & 11:30 AM & 3:30 AM \\
\bottomrule
\end{tabularx}

\end{table}

\begin{table}
    \caption{Cluster statistics}
      \begin{subtable}[t]{\linewidth}
        \centering
        \caption{Distance metrics for the ratio of \harmful{} content when aligned by features of the cluster activity curves. The minimum feature for each metric is indicated in bold font. See \cite{Witowski2012, Frechet1906, Jekel2019, AndradeCampos2012,
        Berndt1994} for details on the individual metrics.}
        \label{tab:similarity_ratio_by_activity}
        \begin{tabularx}{\linewidth}{>{\raggedright\arraybackslash}p{6.5em}*{7}{>{\raggedleft\arraybackslash}X}}
\toprule
 & Partial Curve Mapping
 & discrete Frechet distance
 & area between curves
 & curve length 
 & Dynamic Time Warping
 & mean absolute error 
 & mean squared error \\
\midrule

clock time & 2.9e+01 & \bfseries 9.9e-02 & \bfseries 1.5e+00 & \bfseries 3.4e+00 & \bfseries 6.0e+00 & \bfseries 3.1e-02 & \bfseries 3.0e-03 \\
min activity & 2.4e+01 & 1.1e-01 & 1.5e+00 & 3.5e+00 & 6.2e+00 & 3.3e-02 & 3.3e-03 \\
max activity & 2.9e+01 & 1.3e-01 & 1.7e+00 & 4.1e+00 & 6.9e+00 & 3.6e-02 & 3.7e-03 \\
first inflection & 2.0e+01 & 1.2e-01 & 1.5e+00 & 3.4e+00 & 6.2e+00 & 3.2e-02 & 3.4e-03 \\
first peak & \bfseries 1.8e+01 & 1.1e-01 & 1.6e+00 & 3.6e+00 & 6.4e+00 & 3.3e-02 & 3.3e-03 \\
steepest ascent & 2.0e+01 & 1.2e-01 & 1.5e+00 & 3.4e+00 & 6.2e+00 & 3.2e-02 & 3.4e-03 \\
waking time & 2.0e+01 & 1.1e-01 & 1.6e+00 & 3.5e+00 & 6.3e+00 & 3.3e-02 & 3.2e-03 \\
\bottomrule
\end{tabularx}

      \end{subtable}
    \begin{subtable}[t]{\linewidth}
        \centering
        \caption{Statistics on posts, users and distances per cluster. Distances between and within cluster activity curves. Distances are given using Ward's variance minimization algorithm~\cite{Balcan2014}. The maximum distance within a cluster is indicated in bold font.}
        \label{tab:cluster_stats}
        \begin{tabularx}{\linewidth}{p{5em}*{6}{>{\raggedleft\arraybackslash}X}}
\toprule
 & \multirow{2}{4em}{\centering posts} 
 & \multirow{2}{5em}{users} 
 & \multirow{2}{6em}{posts per user} 
 & \multicolumn{3}{c}{distances} \\
 \cmidrule{5-7}
&&&& morning & intermediate & evening \\
\midrule

infrequent & 7,858,209 & 860,228 & 9 & 2.494 & 2.352 & 2.706 \\
morning & 3,208,484 & 3,599 & 891 & \bfseries 2.494 & 3.324 & 4.258 \\
intermediate & 4,162,911 & 4,297 & 969  & 3.324 & \bfseries 2.353 & 3.956 \\
evening & 2,919,309 & 3,349 & 872 & 4.258 & 3.956 & \bfseries 2.706 \\
bot & 63,313 & 25 & 2,533 &- &- &-\\
\bottomrule
\end{tabularx}
      \end{subtable}
\end{table}

\clearpage

\section*{Supplementary Notes}
\renewcommand{\thesubsection}{Supplementary Note \Alph{subsection}}

\subsection{User Activity Clustering in Germany} \label{sec:suppl:Germany}
To ensure robustness of clustering method and conclusions, we cross-analysed user activity on Twitter originating within Germany. Our dataset encompassed 18,162,387 Tweets, Retweet and Replies authored within the same time span as our main corpus of Tweets originating out of Italy (January 22, 2020, up to August 1st, 2022).

As in the case of Italy, clustering user activity according the the same method resulted in the three distinct clusters. No filtering of suspicious bot-like activity was necessary. The clusters strongly resemble those found within Italy in their activity patterns. Following the same naming convention, the waking times differ from those of their Italian counterparts by an hour at most (\supplref{Germany:cluster_stats}).

\begin{figure}
    \centering
    \caption{Smoothed diurnal activity aligned by waking time, and ratio of \harmful{} content aligned by clock time across clusters in Germany (solid line). For each cluster, the one to two highest peaks of activity and the highest ratio are stressed and annotated with time of occurrence. The dotted curves represent activity and ratio of \harmful{} content in Italy.\label{Germany:comparison_activity_ratio}}
    \begin{subfigure}[b]{\textwidth}
        \centering
        \includegraphics[height=1.2em]{figures_paper/legend_user_types_horizonal}\\
    \end{subfigure}
    \begin{subfigure}[b]{.5\textwidth}
        \includegraphics[width=\columnwidth]{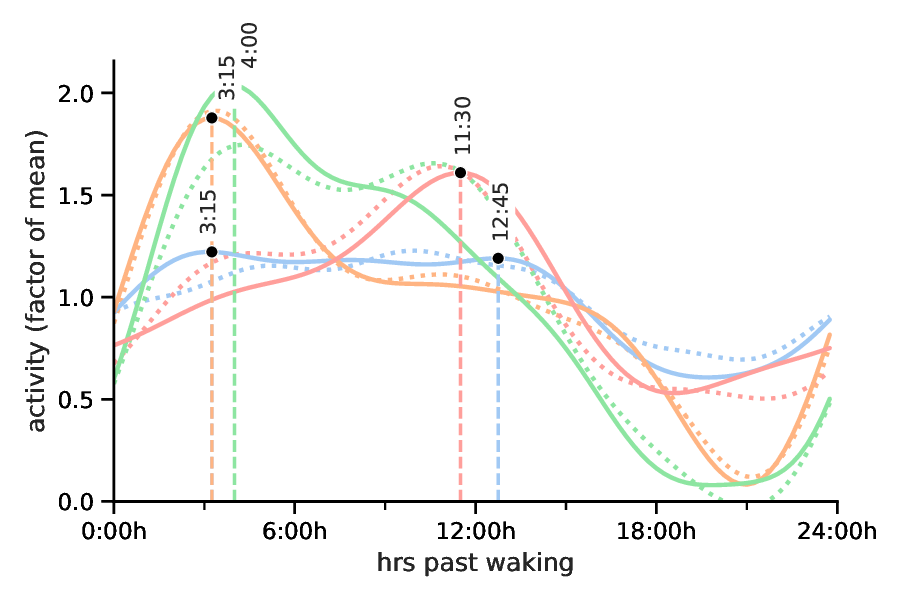}
        \caption{activity by waking time\label{Germany:comparison_activity_waking}}
    \end{subfigure}%
    \begin{subfigure}[b]{.5\textwidth}
        \includegraphics[width=\columnwidth]{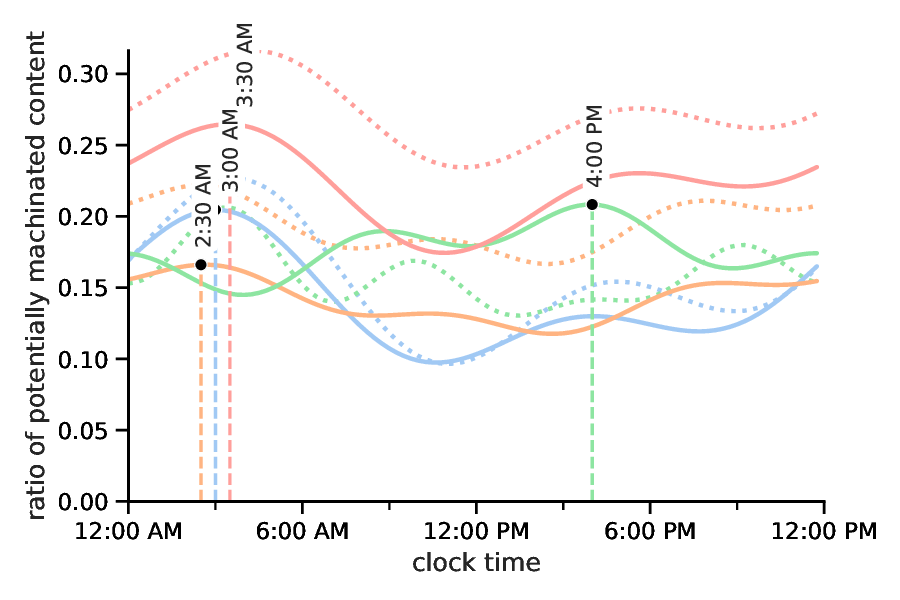} %
        \caption{ratio by time of day\label{Germany:comparison_ratio_clocktime}}
    \end{subfigure}%
\end{figure}

The peaks of activity also fall around the same times as they do for their counterparts, as evident in \supplref{Germany:comparison_activity_waking}.
However, only German \emph{intermediate types} exhibit more than one peak in activity (\supplref{Germany:activity_ratio_stats}).

The ratios of \harmful{} content differ more strongly in between Germany and Italy (\supplref{Germany:comparison_ratio_clocktime}). Morning and evening types in particular spread less \harmful{} content in Germany, although the diurnal patterns are remarkably similar. Only the curve of \harmful{} content ratios of intermediate types appear to follow a different logic. \emph{Intermediate types} display a peak in ratio of \harmful{} content spreading at 4 PM, contrary to the common trend (\supplref{Germany:activity_ratio_stats}).

\begin{table}
    \caption{Statistics on clusters formed by Tweets originating from Germany.}
    \begin{subtable}[t]{\linewidth}
        \caption{Cluster statistics on posts, users, distances, and onset and end of heightened activity per cluster. Distances are given using Ward's variance minimization algorithm~\cite{Balcan2014}, with the maximum distance within each cluster indicated in bold font.}
        \label{Germany:cluster_stats}
        \begin{tabularx}{\linewidth}{p{5em}*{8}{>{\raggedleft\arraybackslash}X}}
\toprule
 & \multicolumn{3}{c}{general} & \multicolumn{3}{c}{distances} & \multicolumn{2}{c}{heightened activity} \\
  \cmidrule(lr){2-4}\cmidrule(lr){5-7}\cmidrule(lr){8-9}
 & posts & users & posts /user & morning & intermediate & evening & onset & end \\
\midrule
infrequent & 8,554,176 & 911,795 & 9 & 2.182 & 1.911 & 2.591 & 8:00 AM & 12:00 AM \\
morning & 3,576,397 & 4,326 & 827 & \bfseries 2.182 & 2.778 & 4.296 & 6:00 AM & 10:00 PM  \\
intermediate & 1,897,997 & 2,521 & 753 & 2.778 & \bfseries 1.860 & 3.838 & 7:45 AM & 11:45 PM \\
evening & 4,133,817 & 4,692 & 881 & 4.296 & 3.838 & \bfseries 2.706 & 10:30 AM & 2:30 AM  \\
\bottomrule
\end{tabularx}
    \end{subtable}
    \begin{subtable}[t]{\linewidth}
        \caption{Times of maximum and minimum activity as well as  of the two highest maximum and lowest minimum ratios per cluster in Germany.}
        \label{Germany:activity_ratio_stats}

\begin{tabularx}{\linewidth}{
>{\raggedright\arraybackslash}p{1.5em} 
>{\raggedright\arraybackslash}p{5em} 
l
Xrr>{\raggedleft\arraybackslash}Xr}
\toprule
& & \multicolumn{3}{c}{max} & \multicolumn{3}{c}{min} \\
  \cmidrule(lr){3-5}\cmidrule(lr){6-8}
& & clock time & hours past waking & activity/ratio & clock time & hours past waking & activity/ratio \\
\midrule
\multirow{8}{*}{\rotatebox[origin=c]{90}{\small activity}}
& \multirow[l]{3}{*}{infrequent} & 11.25 & 3.25 & 0.013 & 3.75 & 19.75 & 0.006 \\
& & 20.75 & 12.75 & 0.012 & 18.25 & 10.25 & 0.012 \\
& & 15.75 & 7.75 & 0.012 & 14.00 & 6.00 & 0.012 \\
\cline{2-2}
& morning & 9.25 & 3.25 & 0.020 & 3.00 & 21.00 & ~0.001 \\
\cline{2-2}
& intermediate & 11.75 & 4.00 & 0.021 & 3.50 & 19.75 & 0.001 \\
\cline{2-2}
& evening & 22.00 & 11.50 & 0.017 & 5.00 & 18.50 & 0.006 \\
\midrule
\multirow{11}{*}{\rotatebox[origin=c]{90}{\small ratio}}
& \multirow[l]{2}{*}{infrequent} & 3.00 & 19.00 & 0.205 & 10.50 & 2.50 & 0.097 \\
& & 16.00 & 8.00 & 0.130 & 19.75 & 11.75 & 0.119 \\
\cline{2-2}
& \multirow[l]{3}{*}{morning} & 2.50 & 20.50 & 0.166 & 14.50 & 8.50 & 0.118 \\
& & 20.50 & 14.50 & 0.153 & 8.50 & 2.50 & 0.130 \\
& & 10.25 & 4.25 & 0.132 & 22.25 & 16.25 & 0.152 \\
\cline{2-2}
& \multirow[l]{3}{*}{intermediate} & 16.00 & 8.25 & 0.208 & 4.00 & 20.25 & 0.145 \\
& & 8.75 & 1.00 & 0.190 & 20.75 & 13.00 & 0.164 \\
& & - & - & - & 11.75 & 4.00 & 0.179 \\
\cline{2-2}
& \multirow[l]{2}{*}{evening} & 3.50 & 17.00 & 0.265 & 11.00 & 0.50 & 0.174 \\
& & 17.50 & 7.00 & 0.230 & 21.25 & 10.75 & 0.221 \\
\bottomrule
\end{tabularx}
    \end{subtable}
\end{table}

\subsection{Behaviour of verified and unverified users} \label{sec:suppl:unverified}
Verified and unverified users exhibit some structural differences in their posting habits. The ratios of \harmful{} posts are significantly different($\chi^2=8801.2$ and $3127.25$ for the raw and Fourier smoothed ratio values, respectively, with both \pvalue$>.001$), with verified users posting higher values of reliable content (\supplref{tab:content_type_ratios_by_verified}).

\begin{figure*}
    \centering
    \caption{Smoothed diurnal activity aligned by waking time, and ratio of \harmful{} content aligned by clock time according to clusters formed from only unverified users (solid line). For each cluster, the one to two highest peaks of activity and ratio are stressed and annotated with time of occurrence. The dotted lines represent activity and ratio of \harmful{} content across clusters formed from users independently of their verification status. \label{unverified:comparison_activity_ratio}}
    \begin{subfigure}[b]{\textwidth}
        \centering
        \includegraphics[height=1.2em]{figures_paper/legend_user_types_horizonal}\\
    \end{subfigure}
    \begin{subfigure}[b]{.5\textwidth}
        \includegraphics[width=\columnwidth]{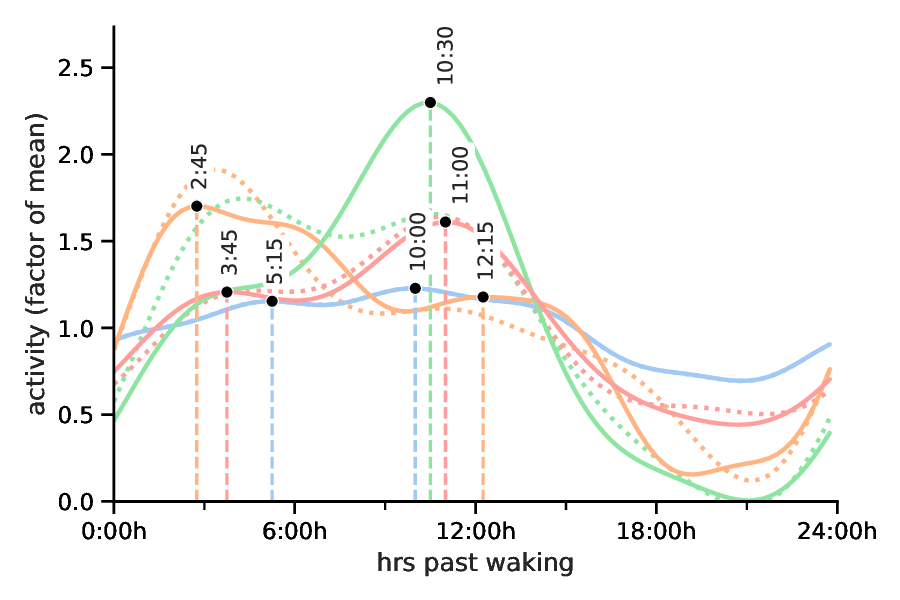} %
        \caption{activity by waking time \label{unverified:comparison_activity_waking_unverified}}
    \end{subfigure}%
    \begin{subfigure}[b]{.5\textwidth}
        \includegraphics[width=\columnwidth]{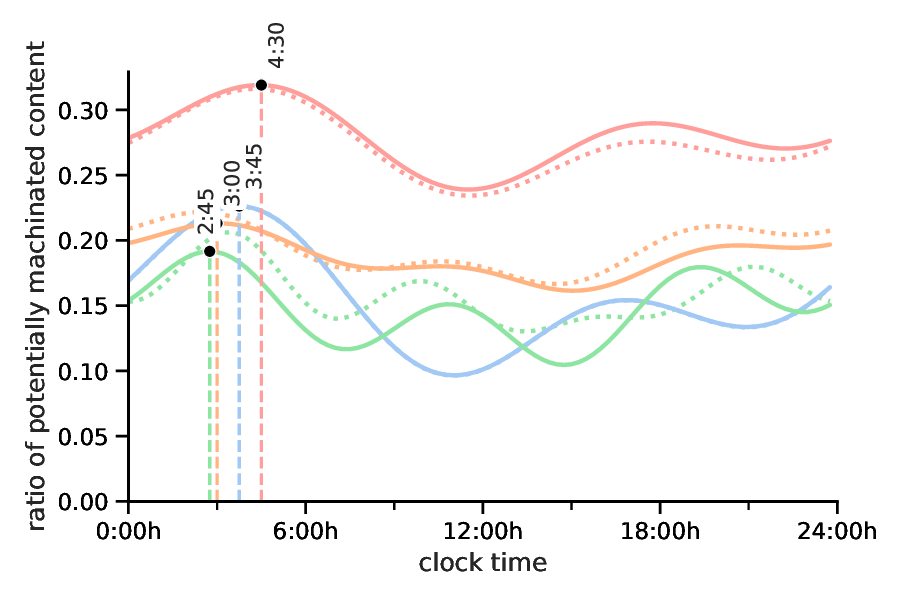} %
        \caption{ratio by time of day\label{unverified:comparison_ratio_clocktime_unverified}}
    \end{subfigure}%
\end{figure*}

Clustering only unverified users results qualitatively similar clusters to those found when clustering independently of verification status (\supplref{unverified:comparison_activity_waking_unverified}). The \emph{\scluster{}} cluster, however, has a pronounced peak at 6:30 pm and no true peak in activity in the morning.
The ratios of \harmful{} content, however, exhibit daily variations remarkably similar to that of the clusters formed from verified as well as unverified users (\supplref{unverified:comparison_ratio_clocktime_unverified}).

\begin{table}[htb]
    \centering
    \caption{Ratios of posts by content type and verification of user.}
    \label{tab:content_type_ratios_by_verified}
    \begin{tabularx}{\textwidth}{Xrrrr}
\toprule
 & \multicolumn{2}{r}{ratio by Tweet} & \multicolumn{2}{r}{ratio by user} \\
\cmidrule(lr){2-3}\cmidrule(lr){4-5}
 & unverified & verified & unverified & verified \\
\midrule
Science & 0.019 & 0.006 & 0.028 & 0.079 \\
Mainstream Media & 0.748 & 0.974 & 0.743 & 0.875 \\
Satire & 0.004 & 0.000 & 0.008 & 0.000 \\
Clickbait & 0.013 & 0.000 & 0.076 & 0.000 \\
Political & 0.102 & 0.013 & 0.107 & 0.038 \\
Fake or hoax & 0.045 & 0.004 & 0.027 & 0.004 \\
Conspiracy \& junk science & 0.068 & 0.003 & 0.012 & 0.003 \\
\midrule \textit{\Harmful{}} & \itshape 0.216 & \itshape 0.020 & \itshape 0.146 & \itshape 0.045 \\
\bottomrule
\end{tabularx}
\end{table}

\end{document}